\documentclass[preprint]{aastex}
\usepackage{color}
\usepackage{natbib,graphicx,latexsym}

\newcommand{\Msini}{\mbox{$M \sin i$}}

\newcommand{\Msun}{\mbox{$M_{\sun}$}}

\newcommand{\Lsun}{\mbox{$L_{\sun}$}}
\newcommand{\Rsun}{\mbox{$R_{\sun}$}}
\newcommand{\Mjup}{\mbox{$M_{\rm Jup}$}}
\newcommand{\Rjup}{\mbox{$R_{\rm Jup}$}}

\newcommand{\ms}{\hbox{m~s$^{-1}$}}
\newcommand{\kms}{\hbox{km~s$^{-1}$}}

\newcommand{\degs}{\mbox{$^{\circ}$}}

\newcommand{\Lbol}{\mbox{$L_{\rm bol}$}}

\newcommand{\Teff}{\mbox{$T_{\rm eff}$}}
\newcommand{\logg}{\mbox{$\log(g)$}}

\shorttitle{Transits and Eclipses in Pan-STARRS-1}
\shortauthors{Dupuy \& Liu}

\begin{document}

\title{Detectability of Transiting Jupiters and Low-Mass Eclipsing
  Binaries in Sparsely Sampled Pan-STARRS-1 Survey Data}

\author{Trent J. Dupuy\altaffilmark{1} and
        Michael C. Liu\altaffilmark{1,2}}

      \altaffiltext{1}{Institute for Astronomy, University of Hawai`i,
        2680 Woodlawn Drive, Honolulu, HI 96822}

      \altaffiltext{2}{Alfred P. Sloan Research Fellow}

\begin{abstract}

  We present detailed simulations of the Pan-STARRS-1 (PS1)
  multi-epoch, multi-band 3$\pi$~Survey in order to assess its
  potential yield of transiting planets and eclipsing binaries.  This
  survey differs from dedicated transit surveys in that it will cover
  the entire Northern sky but provide only sparsely sampled light
  curves.  Since most eclipses would be detected at only a single
  epoch, the 3$\pi$~Survey will be most sensitive to deep eclipses
  ($\gtrsim$~0.10~mag) caused by Jupiters transiting M~dwarfs and
  eclipsing stellar/substellar binaries.  The survey will measure
  parallaxes for the $\sim$4$\times$10$^5$~stars within 100~pc which
  will enable a volume-limited eclipse search, reducing the number of
  astrophysical false positives compared to previous magnitude-limited
  searches.  Using the best available empirical data, we constructed a
  model of the extended solar neighborhood that includes stars, brown
  dwarfs, and a realistic binary population.  We computed the yield of
  deeply eclipsing systems using both a semi-analytic and a full Monte
  Carlo approach.  We examined statistical tests for detecting
  single-epoch eclipses in sparsely sampled data and assessed their
  vulnerability to false positives due to stellar variability.
  Assuming a short-period planet frequency of 0.5\% for M~dwarfs, our
  simulations predict that about a dozen transiting Jupiters around
  low-mass stars ($M_{\star}$~$<$~0.3~\Msun) within 100~pc are
  potentially detectable in the PS1 3$\pi$~Survey, along with
  $\sim$300 low-mass eclipsing binaries (both component masses
  $<$~0.5~\Msun), including $\sim$10 eclipsing field brown dwarfs.
  Extensive follow-up observations would be required to characterize
  these candidate eclipsing systems, thereby enabling comprehensive
  tests of structural models and novel insights into the planetary
  architecture of low-mass stars.

\end{abstract}


\section{Introduction}

Eclipsing systems are the gold standard by which theoretical models of
stars, brown dwarfs, and extrasolar planets are tested, as they enable
direct mass, radius, and temperature measurements. For example, mass
and radius measurements of stars have provided critical tests of
stellar interiors \citep{1991A&ARv...3...91A} and key insights into
stellar magnetism \citep{2007ApJ...660..732L}. In recent years, the
discovery of transiting extrasolar planets has extended such
fundamental measurements to the study of giant planets orbiting nearby
stars \citep[e.g.,][]{2000ApJ...529L..45C}. The existence of such a
population of short-period giant planets was originally a surprise,
and the ongoing measurement of masses and radii for these planets
continues to reveal remarkable diversity \citep[][and references
therein]{2007prpl.conf..701C}, spurring active research to
characterize their atmospheres and understand their formation history.

The first multi-epoch all-sky surveys, such as Pan-STARRS
\citep{2002SPIE.4836..154K}, SkyMapper \citep{2007PASA...24....1K},
and the planned Large Synoptic Survey Telescope
\citep[LSST;][]{2002SPIE.4836...10T}, will soon open the door to the
discovery of eclipsing systems different from those previously
accessible.  All-sky surveys contain the most comprehensive
information for objects in the immediate solar neighborhood, which is
populated predominately with M~dwarfs. Historically, M~dwarfs have
been neglected by surveys for eclipsing binaries and transiting
planets.  This is because M~dwarfs represent a tiny fraction of
sources detected in magnitude-limited stellar variability surveys
\citep[$\sim$1:1000, e.g.,][]{2007ASPC..362...15S}.  In addition,
signal-to-noise (S/N) constraints have limited radial velocity (RV)
searches for extrasolar planets to monitor only the nearest M~dwarfs,
which typically comprise $\sim$10\% of RV samples
\citep[e.g.,][]{2007ApJ...670..833J}.  Consequently, there are only
seven known eclipsing binaries with both component masses
$<$~0.5~\Msun\ \citep{1977ApJ...218..444L, 1999A&A...341L..63D,
  2004A&A...426..577M, 2005ApJ...625L.127C, 2006AJ....131..555H,
  2008ApJ...684..635B, 2009ApJ...701.1436I} and only one extrasolar
planet known to transit an M~dwarf
\citep[Gl~436b,][]{2004ApJ...617..580B, 2007A&A...472L..13G}.

Because of their small size (0.1--0.5~\Rsun), M~dwarfs experience more
extreme eclipses for a given size companion than solar-type stars. For
example, the eclipse depth due to a transiting Jupiter-size object is
$\sim$10\% for a mid-M~dwarf ($\sim$0.3~\Rsun), while it is an order
of magnitude smaller ($\sim$1\%) for a solar-type star. Such deep
eclipses are in principle readily detectable in sparsely sampled light
curves, where the interval between observing epochs is large compared
to the periodic eclipse signal. The power of this approach has been
demonstrated by the discovery of short-period M~dwarf eclipsing
binaries in both the Two Micron All Sky Survey (2MASS) calibration
fields \citep{2008ApJS..175..191P} and the SDSS-II Supernova Survey
\citep{2008ApJ...684..635B}.

Pan-STARRS-1 (PS1) is the first of the large, multi-epoch surveys to
be on sky, having begun extensive on-sky commissioning runs in spring
2009. It will conduct a number of different surveys, including a
conventional, dedicated transit survey pointed at a fixed position on
the sky (Pan-Planets). We have instead focused on the 3$\pi$~Survey
(described in detail in Appendix~\ref{app:ps1}), the largest PS1
survey, which will image the entire 3$\pi$ steradians of sky visible
from Haleakala, on the island of Maui, at 12~epochs spanning 3~years
in each of 5~bandpasses.\footnote{The next largest PS1 survey is the
  Medium Deep Survey, which will image 10 PS1 fields (7 deg$^2$ each)
  at 72~epochs over 3~years in each of 5~bandpasses, with a
  $\sim$4-day cadence designed to detect supernovae. Since it covers
  only $\approx$0.2\% of the area of the 3$\pi$~Survey, the Medium
  Deep Survey is a negligible source of nearby, low-mass eclipsing
  systems.} The cadence is designed to both detect asteroids and
measure the parallaxes of nearby stars and is not specifically
tailored to detect eclipsing binaries and transiting planets. However,
its sparsely sampled light curves will be amenable to the detection of
deep eclipses ($\gtrsim$~0.10~mag) such as those produced by eclipsing
binaries or Jovian planets transiting M~dwarfs. The light curves for
the hundreds of thousands of low-mass stars in the immediate solar
neighborhood with parallaxes measured by PS1 will enable a novel,
volume-limited variability survey of the entire northern sky. Such an
approach is conceptually similar to the MEarth Project
\citep{2008PASP..120..317N, 2009IAUS..253...37I}, which is monitoring
a carefully selected sample of 2000 nearby M~dwarfs for planetary
transits and has already discovered and characterized one eclipsing
binary \citep{2009ApJ...701.1436I}.

Previous transit surveys have been magnitude-limited and have thus
encountered two challenges: (1) overcoming a broad range of
astrophysical false positives that mimic planetary transits and are
more numerous by $\sim$20:1 \citep[e.g.,][]{2005A&A...431.1105B,
  2005A&A...438.1123P, 2006ApJ...644.1237O}; (2) obtaining follow-up
observations on transit host stars that are fainter than those
discovered by dedicated RV surveys. A volume-limited sample naturally
reduces such rare astrophysical false positives and focuses on only
the nearest, and thus brightest, stars in the sky. Also, even if no
events are detected, such a well-defined sample can enable meaningful
statements about the underlying population of eclipsing systems, which
can be challenging to quantify for magnitude-limited samples. However,
sparsely sampled light curves are potentially more vulnerable to
mundane false positives (e.g., due to star spots or photometric
outliers) since by definition they contain no information about the
light curve shape during eclipse.

Previous authors have estimated the transit yield of Pan-STARRS,
considering all stars detected within the magnitude limits of selected
surveys. \citet{2007ASPC..366..273G} and \citet{2008ApJ...686.1302B}
used analytic relations to derive the number of transits for
solar-type stars ($\sim$20) and M~dwarfs ($\sim$40) in the
magnitude-limited sample of the PS1 Medium Deep Survey.
\citet{2008AIPC.1082..275F} used a semi-analytic approach and,
assuming that a large fraction (10\%) of M~dwarfs host a short-period
(1~$<$~$P$~$<$~3~days) Neptune-mass planet, estimated that $\sim$300
transits in the 3$\pi$~Survey and $\sim$10 transits in the Medium Deep
Survey could be detected.  Neither of these results are directly
comparable to our simulations, as we have used substantially different
input assumptions, most importantly restricting ourselves to a
volume-limited sample, and we also evaluate the detectability of such
events in simulated survey light curves.  In addition, we consider
low-mass eclipsing binaries, which produce essentially the same
eclipse signal in the sparsely sampled light curves as Jupiters
transiting M~dwarfs.  Such ``false positives'' in the search for
planets would be interesting in their own right, given the small
number of low-mass eclipsing binaries currently known.

In this paper, we predict the yield of eclipsing binaries and
transiting Jupiters in PS1 3$\pi$~Survey data.  Using the best
available empirical information, we have constructed a model of the
solar neighborhood encompassing solar-type stars to the coldest known
brown dwarfs, including a realistic binary population. We have
developed a method for analytically predicting the planetary transit
yield given an inventory of survey targets.  We also constructed Monte
Carlo simulations to predict the yield of both planets and eclipsing
binaries and to characterize their host stars, while carefully
considering the false positive rate due to star spots.  The prospects
of characterizing the candidate eclipsing systems found in sparsely
sampled light curves with follow-up observations is discussed in
detail, along with the impact of our input assumptions on the survey
yield.


\section{Model of the Solar Neighborhood\label{sec:model}}

We first created a model that describes the population of stellar and
substellar objects within the solar neighborhood.  These are the host
stars for the transiting/eclipsing systems. The distance limit is set
by the astrometric precision of the survey, since this determines the
smallest parallaxes it will be able to measure.  Both PS1 and LSST
have astrometric precision goals of 10~mas per epoch, which will
enable S/N~$>$~10 parallax measurements for objects within 100~pc.
Since this distance is smaller than the scale length and height of the
Galactic disk \citep[2600~pc and 300~pc;][]{2008ApJ...673..864J}, we
assumed that objects are uniformly distributed within this volume.

We assigned the properties of these simulated objects as follows:
(1)~masses and ages were drawn from appropriate initial mass and birth
rate functions; (2)~luminosities and radii were determined from
evolutionary models; (3)~spectral types were assigned based on an
empirical relation with luminosity, (4)~optical and near-infrared
colors and bolometric corrections were assigned based on empirical
relations with spectral type; (5)~binary companions were assigned to
an appropriate fraction of simulated objects, with the orbital
separations and mass ratios randomly drawn from empirical
distributions; and (6)~short-period ($P$~$<$~11.5~days) Jupiter-like
planets were assigned to an appropriate fraction of simulated objects,
with orbital periods randomly drawn from an empirical distribution.

The resulting population of low-mass stars ($M$~$<$~1.0~\Msun) and
brown dwarfs is summarized in Table~\ref{tbl:100pc}.  We list both the
total number of objects within 100~pc (over 4$\pi$ steradians) and the
objects expected to be within the magnitude limits and areal coverage
of the PS1 3$\pi$~Survey, which is described in
Appendix~\ref{app:ps1}.  For example, our model predicts that a total
of 94 M~dwarfs within 100~pc host planets with orbits aligned such
that transits are observable from Earth.  Of the 70 of these in the
3$\pi$ steradians observable by PS1, 56 are expected to be within the
magnitude limits of the 3$\pi$~Survey (14 are saturated).  In our
simulations, the number of actual transits observed by PS1 will be
some subset of these 56, depending on if transit events occur at the
observation epochs.

\subsection{Low-mass Stars and Brown Dwarfs\label{sec:sim-star}}

As the starting point for our simulation, we drew masses and ages of
objects from appropriate random distributions.  For ages, we assumed a
constant star formation rate over a period of 10~Gyr
\citep[e.g.,][]{2002AJ....123.3356G, 2003A&A...409..523R}. For the
mass function, we used the best fit log-normal function derived by
\citet{2008AJ....136.1778C}.  The number of objects in our simulation
was determined by the normalization of this mass function
(0.1~$\log(M/\Msun)^{-1}$~pc$^{-3}$), which predicts that there are
4.0$\times$10$^5$ objects within 100~pc.


To determine the physical properties of the simulated objects, we used
solar-metallicity Lyon evolutionary models \citep{1998A&A...337..403B,
  2000ApJ...542..464C, 2003A&A...402..701B} to derive luminosities and
radii. We created a 2-dimensional array, uniformly gridded in
$\log{M}$ and $\log(\rm age)$, for each model-predicted property.
Then, for each simulated object, we bilinearly interpolated its
model-predicted $\log\Lbol$ and $\log{R}$ from these arrays. Because
our simulated objects span a wide range of masses and temperatures
(from low-mass main-sequence stars to old brown dwarfs), different
evolutionary models are appropriate for different simulated objects.
We used the models of \citet{1998A&A...337..403B} for objects where
these models predicted an effective temperature $>$~2800~K, i.e., too
hot to harbor appreciable amounts of dust in the atmosphere
\citep[e.g.,][]{1996A&A...308L..29T, 1997ApJ...480L..39J}. We used the
DUSTY models of \citet{2000ApJ...542..464C} for objects where these
models predicted effective temperatures $>$~1400~K and the COND models
of \citet{2003A&A...402..701B} for the remaining objects. This break
in temperature roughly corresponds to the transition from L to T
dwarfs \citep[e.g.,][]{gol04, me-2397a}, where dust clouds sediment
below the photosphere.

\begin{figure} 
\centerline{\includegraphics[width=6.5in,angle=0]{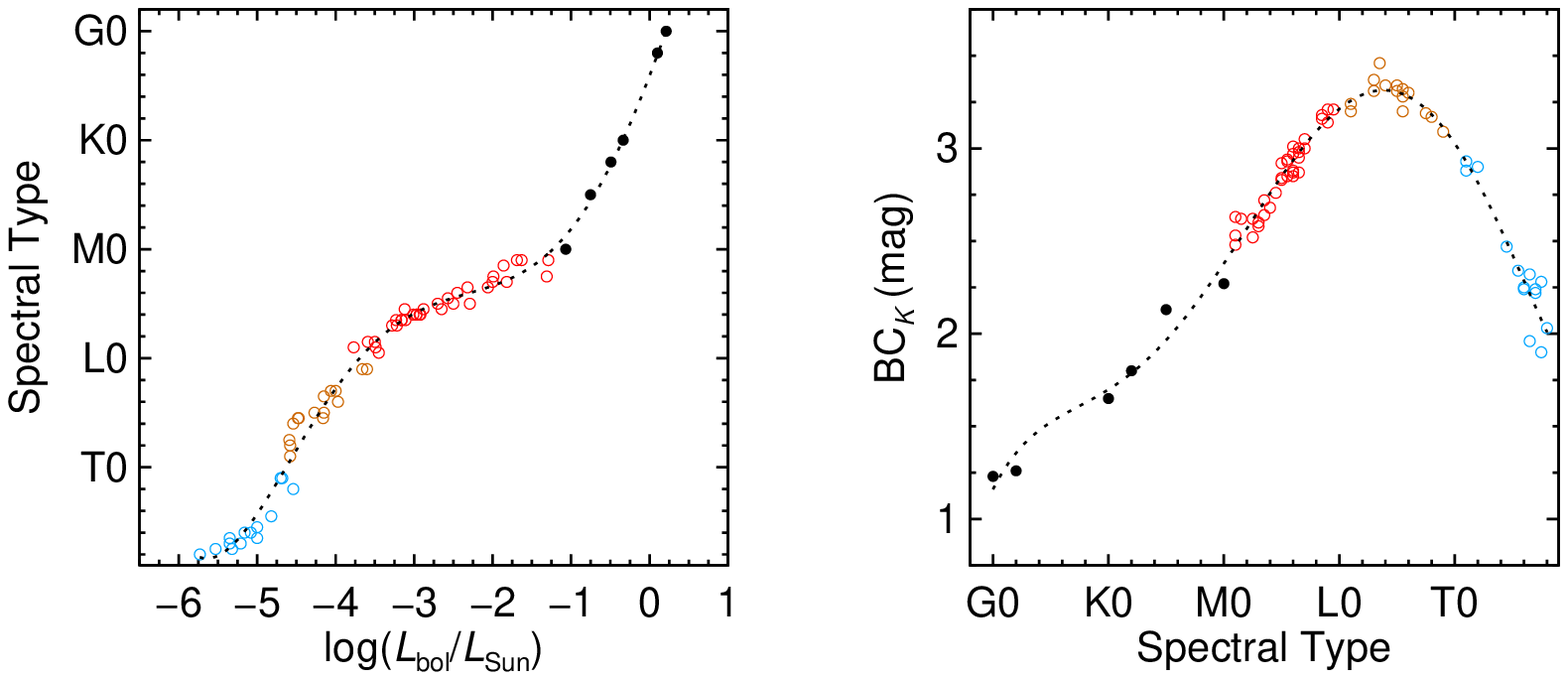}}

\caption{ \normalsize Empirical relations between spectral type,
  luminosity, and $K$-band bolometric correction.  Open circles are
  objects with measured spectral types and luminosities, with M, L,
  and T dwarfs show in red, brown, and blue, respectively
  \citep{2000ApJ...535..965L, leg01, gol04}.  Filled black circles
  represent tabulated values from \citet{2000asqu.book.....C}.  The
  dotted lines show our fifth-order polynomial fits to these data,
  which we used to assign spectral types and BC$_K$ values to objects
  in our simulations (see Table~\ref{tbl:spt-rel} for the
  coefficients). \label{fig:spt-rel}}

\end{figure}

We assigned spectral types to the simulated objects based on their
model-predicted luminosities.  We compiled the luminosities derived by
\citet{2000ApJ...535..965L, leg01} and \citet{gol04} for M, L, and T
dwarfs with measured distances and spectral energy distributions.  We
were careful to exclude known and suspected binaries from this sample
\citep{2006astro.ph..5037L}.  This resulted in 63 objects ranging in
spectral type from M1 to T8, adopting near-infrared spectral types
when possible.  We supplemented this sample with luminosities for the
spectral types G0, G2, K0, K2, K5, and M0 tabulated by
\citet{2000asqu.book.....C}, and originating from
\citet{schmidt-kaler1982} and \citet{1987A&A...177..217D}.  In order
to assign spectral types to simulated objects, we derived a polynomial
relation between spectral type (SpT)\footnote{In assigning numerical
  spectral types, we assumed each spectral class comprised 10
  subclasses, even though this is not always formally correct (e.g.,
  for the K spectral class).} and \Lbol, which is shown in
Figure~\ref{fig:spt-rel} with the polynomial coefficients given in
Table~\ref{tbl:spt-rel}.

The last remaining property that we determined for simulated objects
were their absolute magnitudes in the relevant optical and
near-infrared bandpasses.  We chose not to use the absolute magnitudes
predicted by evolutionary models because it is well known that they do
not accurately reproduce the colors of very low-mass stars and brown
dwarfs \citep[e.g.,][]{2009ApJ...692..729D, me-2397a}.  Instead we
used $K$-band bolometric corrections (BC$_K$) combined with observed
optical and near-infrared colors.  \citet{2000ApJ...535..965L, leg01}
and \citet{gol04} derived BC$_K$ values for all 63 M, L, and T dwarfs
with measured luminosities.  We again supplemented this sample with
tabulated values from \citet{2000asqu.book.....C}, which are based on
$V$-band bolometric corrections from \citet{1966ARA&A...4..193J} and
$V-K$ colors from \citet{1988PASP..100.1134B} and
\citet{1991AJ....101..662B}.  We fit a polynomial relation between
BC$_K$ and SpT in order to determine bolometric corrections, and thus
absolute magnitudes in $K$-band, for simulated objects.  The fit is
shown in Figure~\ref{fig:spt-rel}, and the coefficients are given in
Table~\ref{tbl:spt-rel}.

\begin{figure} 
\centerline{\includegraphics[width=3.5in,angle=0]{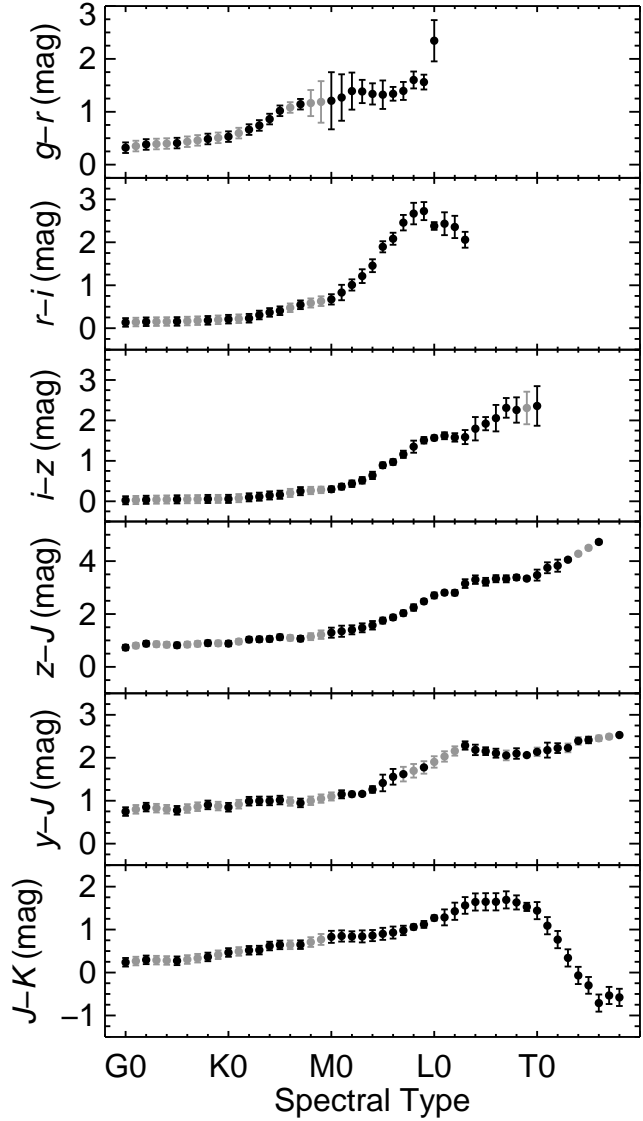}}

\caption{ \normalsize Optical and near-infrared colors of low-mass
  stars and brown dwarfs compiled from the literature or computed by
  us from synthetic photometry, as described in the
  Section~\ref{sec:sim-star}.  The filled circles and error bars
  represent the mean color at each spectral type and the rms.  If
  fewer than 2 objects were available at a given spectral type, the
  mean was interpolated from the neighboring bin, likewise 3 objects
  were required for the rms.  Gray symbols indicate spectral types
  where the mean and/or rms were interpolated. The color of each
  object in our simulation was drawn from a Gaussian distribution with
  a mean and rms appropriate for its spectral
  type. \label{fig:spt-color}}

\end{figure}

To determine the optical and near-infrared colors of the low-mass
stars in our simulation, we compiled existing photometry and also
computed colors from synthetic photometry of publicly available
spectra (Appendix~\ref{app:ps1}).  To properly account for the
observed intrinsic scatter in color at each integer spectral type, we
drew colors from Gaussian distributions with the observed mean and
root-mean-squared (rms).  For M0--L0 dwarfs, we used the tabulated
mean and rms $g-r$ colors from \citet{2007AJ....133..531B} and the
tabulated mean and rms $r-i$, $i-z$, $z-J$, and $J-K$ colors from
\citet{2008AJ....135..785W}.\footnote{No $g$-band photometry or
  spectroscopy is available for spectral types later than L0, so we
  cannot assign $g-r$ colors beyond L0, implicitly assuming they will
  be undetectable in $g$-band for the purposes of this simulation.}
For L1--T8 dwarfs, we used the $r-i$, $i-z$, and $z-J$ colors of the
196 objects from \citet{2002astro.ph..4065H},
\citet{2004AJ....127.3553K}, and \citet{chiu05}, after excluding
objects known or suspected to be binaries.  In Appendix~\ref{app:ps1},
we have derived conversions between the SDSS and PS1 photometric
systems, which we applied to this SDSS photometry.\footnote{Since
  there are no SDSS measurements in $g$ and $r$ band for later
  spectral type objects than we have derived photometric conversions
  for ($>$~M6 in $g$, $>$~L5 in $r$), we simply use the conversion at
  the most extreme spectral type for which it is defined at all later
  spectral types.  This affects very few objects as such faint, red
  objects are not typically detected in the bluest bandpasses.} For
the $J-K$ colors of L1--T8 dwarfs, we used the tabulated mean and rms
values of \citet{2009AJ....137....1F}, converting these 2MASS colors
to the MKO photometric system using the relations of
\citet{2004PASP..116....9S}. For G0--K7 dwarfs, we used the
\citet{1998PASP..110..863P} library of averaged stellar spectra to
derive colors from synthetic photometry, adopting 0.10~mag for the
intrinsic rms scatter at these spectral types. Finally, because the
$y$ bandpass for PS1 is unique, we derived $y-J$ colors solely from
synthetic photometry of the publicly available spectra described in
Appendix~\ref{app:ps1}. The compiled and derived colors are shown as a
function of spectral type in Figure~\ref{fig:spt-color} and
Table~\ref{tbl:spt-color}. We required a minimum of 2 objects at a
given spectral type to compute the mean, and 3 objects to compute the
rms. Where the mean or rms was not available for a given spectral
type, we interpolated its value from the neighboring spectral type
bins.

\subsection{Binary Companions\label{sec:sim-comp}}

We used the results of multiplicity studies targeting specific
mass/spectral type intervals of low-mass stars and brown dwarfs to
assign binary companions to simulated objects. 

\begin{figure} 
\centerline{\includegraphics[width=6.5in,angle=0]{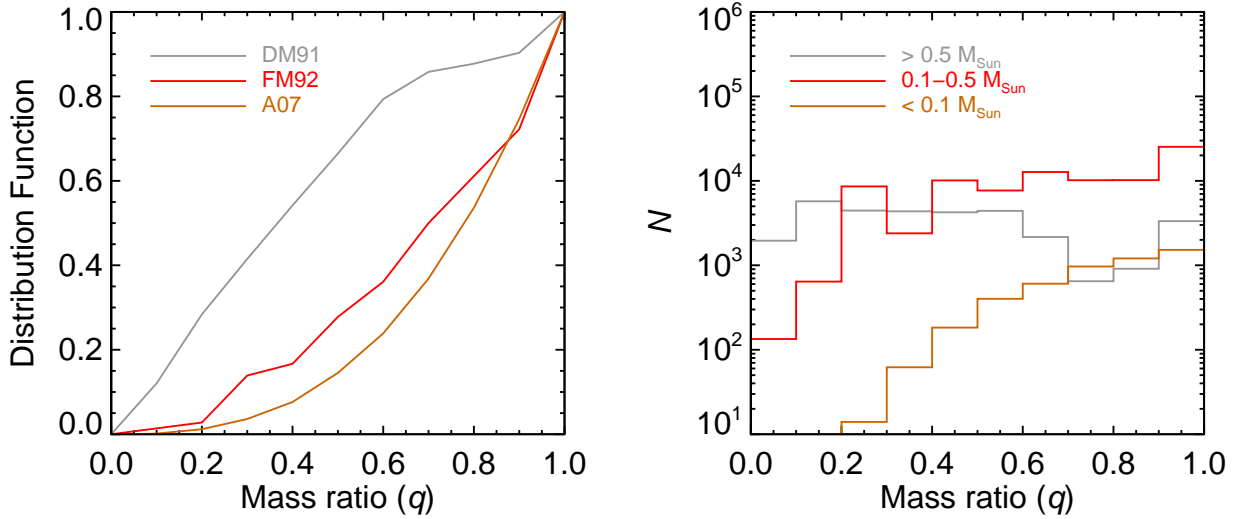}}

\caption{ \normalsize Mass ratio distribution of simulated binary
  companions in the PS1 3$\pi$~Survey. We drew mass ratios from the
  cumulative distribution functions of objects from DM91 and FM92 for
  FGK stars ($>$~0.5~\Msun) and early to mid-M~dwarfs
  (0.1--0.5~\Msun), respectively.  For very low-mass stars and brown
  dwarfs ($<$~0.1~\Msun), we used the power law of
  \citet{2007ApJ...668..492A}. \label{fig:qrat}}

\end{figure}

For objects with masses $>$~0.5~\Msun\ (i.e., FGK dwarfs), we adopted
the results of \citet[][DM91 hereinafter]{1991A&A...248..485D}. We
drew mass ratios ($q$~$\equiv$~$M_2/M_1$) according to the DM91
cumulative distribution for $P$~$<$~10$^4$~day binaries corrected for
viewing angle effects and completeness (i.e., line 3 of their Table
7).\footnote{We caution that the lowest mass ratio bins of DM91
  ($q$~$<$~0.2) depend strongly on their assumed completeness
  corrections as they were not sensitive to very low-mass
  companions. However, this has no impact on our results as we found
  that all $>$~0.5~\Msun\ stars within 100~pc are saturated in the PS1
  3$\pi$ Survey.} This cumulative distribution and the resulting mass
ratio distribution are shown in Figure~\ref{fig:qrat}. We drew periods
from a log-normal distribution with a mean of $\log(P/{\rm
  days})$~=~4.8 and $\sigma_{\log{P}}$~=~2.3 and used the system mass
$M(1+q)$ to convert the orbital period to semimajor axis ($a$). We
assigned companions to 57\% of $M$~$>$~0.5~\Msun\ objects, in
accordance with the binary frequency found by DM91.

For objects with masses 0.1~$<$~$M$~$<$~0.5~\Msun (roughly M0--M6
dwarfs), we adopted the results of \citet[][FM92
hereinafter]{1992ApJ...396..178F}.  They found a log-normal
distribution of orbital periods consistent with that of DM91, which
implies somewhat more compact orbits than for FGK dwarf binaries at a
given period according to Kepler's Third Law ($a \propto M^{1/3}$).
We drew mass ratios from the cumulative distribution of the entire
sample of FM92 (their Table~1).  This cumulative distribution and the
resulting mass ratio distribution are shown in Figure~\ref{fig:qrat}.
Companions were assigned to 42\% of simulated objects in this mass
range, in accordance with the binary frequency found by FM92.

The binary properties of very low-mass stars and brown dwarfs
($M$~$<$~0.1~\Msun) are less well constrained by observations than for
FGKM~dwarfs. However, two attributes of very low-mass binaries have
become clear from previous multiplicity studies: (1) binaries with
separations wider than $\gtrsim$~15~AU are rare
\citep[e.g.,][]{2003ApJ...587..407C} and (2) mass ratios are rarely
very different from unity \citep[e.g.,][]{2007prpl.conf..427B}.  From
a uniform analysis of the results from high-resolution imaging surveys
targeting $>$~M6 dwarfs, \citet{2007ApJ...668..492A} found that the
mass ratio distribution of binaries was adequately described by a
power law ($dN/dq$~=~$q^{-\gamma}$), where
$\gamma$~=~$-$1.8$\pm$0.6.\footnote{This is somewhat at odds with the
  power-law slope of $\gamma$~=~$-$4.2$\pm$1.0 derived by
  \citet{2007prpl.conf..427B} for L and T dwarfs.
  \citet{2007ApJ...668..492A} suggests that this is due to a
  steepening of the power law at lower masses, since the
  \citet{2007prpl.conf..427B} sample focuses on lower mass
  objects. Regardless, the mass ratio distribution has a negligible
  effect on our primary result (i.e., the survey yield of eclipsing
  systems) since for very low-mass stars and brown dwarfs the
  mass--radius relation is essentially flat.}  We used this power law
to draw the mass ratios of companions, and the resulting distribution
is shown in Figure~\ref{fig:qrat}.  Imaging surveys have been able to
constrain the distribution of orbital separations outside of
$\sim$2~AU, and \citet{2003ApJ...587..407C} found that 15\% of $>$~M8
dwarfs have companions at wider separations than this.  Accounting for
the detection limits of existing surveys, \citet{2007ApJ...668..492A}
suggested an additional $\sim$5\% of objects have undetected
companions with separations $\gtrsim$~1~AU.  The only existing surveys
sensitive to the much tighter separations relevant to eclipsing
systems are RV surveys \citep[e.g.,][]{2003A&A...401..677G,
  2006A&A...448..655J, 2006AJ....132..663B}.
\citet{2005MNRAS.362L..45M} found that a significant fraction of
objects (an additional 20--30\%) may harbor close companions, given
four spectroscopic binaries identified in RV surveys.
\citet{2006AJ....132..663B} contributed additional radial velocity
data and, using a slightly different analysis, found that the fraction
of objects hiding close companions is 11$^{+7}_{-4}$\%.  In fact, they
found that a log-normal period distribution could reasonably explain
the ensemble of data if the distribution was truncated at 15~AU.  We
drew orbital periods from this truncated log-normal distribution, with
a peak at $\log(P/{\rm days})$~=~3.8 and $\sigma_{\log{P}}$~=~1.3.
Companions were assigned to 26\% (i.e., $15+11$\%) of
$M$~$<$~0.1~\Msun\ objects.

The physical and observational properties of the companions were
determined in the same manner as described for single stars in
Section~\ref{sec:sim-star}. Companions falling outside the luminosity
range of G0--T8 dwarfs were excluded from the simulation, and for the
remaining companions we included their flux when considering survey
detectability. The short-period tail of the log-normal distributions
we used to assign companions resulted in some systems that would be
non-detached binaries. Since we are interested in detached eclipsing
systems, we excluded binary companions where the radius of the primary
intruded into the companion's Hill sphere
($R_1~>~a(1~-~(\frac{M_2}{3M_1})^{1/3})$).

\subsection{Short-period Jupiters\label{sec:sim-plnt}}

For solar-type host stars, the population of Jupiter-like planets in
short-period orbits is very well characterized, despite their rarity,
thanks to long-term radial velocity observing programs that have
targeted thousands of stars. In a recent analysis of the Keck Planet
Search, \citet{2008PASP..120..531C} found that 1.1$\pm$0.5\% of
targets have a massive planet (\Msini~$>$~0.5~\Mjup) in a short-period
orbit ($P$~$<$~11.5~days).\footnote{As discussed in Section~2.2 of
  \citet{2008ApJ...686.1302B}, the bias of RV surveys
  toward metal-rich stars may result in an overestimatation of the
  planet frequency by as much as a factor of $\sim$2, in comparison to
  a volume-limited sample.}

Fewer planets are known around M~dwarfs, so their population of
short-period planets is not as well characterized. Only one transiting
planet around an M~dwarf is known
\citep[Gl~436b,][]{2007A&A...472L..13G}, with a mass of 0.072~\Mjup\
and radius of 0.44~\Rjup\ \citep{2007ApJ...671L..65T}. Like many of
the other known planets around M~dwarfs, Gl~436b is much less massive
than Jupiter and more akin to Neptune in its mass and radius.
\citet{2007ApJ...670..833J} used the California and Carnegie Planet
Search detections of giant planets (\Msini~$>$~0.8~\Mjup) within
2.5~AU to derive a giant planet fraction of 4.2$\pm$0.7\% for FGK
stars (0.7~$<$~$M_{\star}$~$<$~1.3~\Msun) and 1.8$\pm$1.0\% for late-K
and M dwarfs ($M_{\star}$~$<$~0.7~\Msun). We adopt the ratio of these
planet fractions as the relative rate of occurence of planets around
M~dwarfs compared to solar-type stars. Thus, if 1.1\% of solar-type
stars have short-period Jupiters \citep{2008PASP..120..531C}, then
0.47\% of M dwarfs are assumed to have such planets.

We assigned a short-period planet to 0.47\% of simulated objects with
$M_{\star}$~$<$~0.7~\Msun and to 1.1\% of simulated objects with
$M_{\star}$~$>$~0.7~\Msun. We drew orbital periods from the cumulative
distribution of all short-period planets ($P$~$<$~11.5~days) in the
Catalog of Nearby Exoplanets
\citep{2006ApJ...646..505B}.\footnote{\url{\texttt{http://exoplanets.org/planets.shtml}}}
The median orbital period was 3.8~days, and the mean and rms were
4.5$\pm$2.3~days.  We assumed a single radius of 1.0~\Rjup\ for all
planets and that the planets contributed no flux to the system. We
also assumed circular orbits, as short-period orbits are expected to
be tidally circularized, and neglected the mass of the planet when
converting the orbital period to semimajor axis.


\section{Survey Yield Predictions\label{sec:sims}}

We have taken two approaches to simulating the yield of transit events
in sparsely sampled multi-epoch survey data: (1) a semi-analytic
approach which assigns transit detection probabilities to individual
survey stars and (2) a full Monte Carlo simulation in which light
curves for all survey stars are generated and searched for transit
events. In both cases, we assume that survey light curves will be
randomly sampled in phase, neglecting any window function
effects. This assumption is valid for surveys in which the typical
interval between observation epochs is much longer than that of the
periodic signal being sought.

Sparsely sampled light curves require a different approach to eclipse
detection than is used for the densely sampled light curves from
conventional transit surveys. The most important difference is that
transits are expected to only be detected at a single
epoch. Appendix~\ref{app:alg} describes the eclipse detection
algorithms we have considered. For the simulations described below, we
used a ``S/N model'' (algorithm 3 in Appendix~\ref{app:alg}) to assess
the transit yield. Our results do not depend strongly on the algorithm
used because they all have comparable detection sensitivities. For
real data, the best algorithm will depend on the number of light curve
points (Figure~\ref{fig:alg-mindepth}) and the dominant source of
false positives (Figure~\ref{fig:alg-false-pos}).

\subsection{Semi-analytic Calculation\label{sec:semianal}}

We first consider an analytic method to assign transit detection
probabilities to survey stars. The following approach to computing the
survey yield is only semi-analytic because it uses as a starting point
the Monte Carlo simulation of the solar neighborhood described in
Section~\ref{sec:model}. However, given real survey data for a
well-characterized set of target stars this approach could be used to
compute the expected transit yield directly, given assumptions about
the underlying planet population.

The concept behind this approach is that each survey star has some
mininum detectable fractional eclipse depth ($\delta_{\rm min}$),
given the photometric S/N of the star (S/N$_{\star}$), the total
number of points in its light curve ($n$), the number of points in
which the transit was observed ($n_{\rm obs}$), and the transit S/N
threshold (S/N$_{\rm min}$).  We consider two cases, one in which the
photometric S/N of the star is photon-limited and improves as
$\sqrt{n}$, and one in which it is at the systematic noise floor
(S/N$_{\rm floor}$), which for PS1 is expected to be about 100 (i.e.,
$\sigma_{\rm floor}/N_{\star}$ = 1\%).  To derive $\delta_{\rm min}$,
we first use Equation~\ref{eqn:sntr} to find how the transit S/N
depends on the fractional eclipse depth ($\delta$):
\begin{equation} \label{eqn:sneq} {\rm S/N}_{\rm transit} = 
  \left\{\begin{array}{cccc}
      \frac{{N_{\star}}\delta}{\sqrt{\frac{N_{\star}}{n-n_{\rm obs}} + \frac{N_{\star}(1-\delta)}{n_{\rm obs}}}} & = &
      \frac{{\rm S/N}_{\star}\delta}{\sqrt{\frac{1}{n-n_{\rm obs}} + \frac{1-\delta}{n_{\rm obs}}}}  &  
      {\rm if~S/N}_{\star} < {\rm S/N}_{\rm floor} \\
 & & & \\
      \frac{{N_{\star}}\delta}{\sqrt{N_{\star}^2\sigma_{\rm floor}^2 + N_{\star}^2(1-\delta)^2\sigma_{\rm floor}^2}} & = &
      \frac{{\rm S/N}_{\star}\delta}{\sqrt{1 + (1-\delta)^2}}  &  {\rm if~S/N}_{\star} = {\rm S/N}_{\rm floor} \\
\end{array}\right.
\end{equation}
We can then solve for the eclipse depth corresponding to the transit
detection threshold (S/N$_{\rm min}$):
\begin{equation} \label{eqn:dmin}
\delta_{\rm min} = 
\left\{\begin{array}{lc}
\frac{({\rm S/N}_{\rm min})^2}{2n_{\rm obs}({\rm S/N}_{\star})^2}\left[ \sqrt{1 + \frac{4({\rm S/N}_{\star})^2}{({\rm S/N}_{\rm min})^2} \left( \frac{n_{\rm obs}^2}{n-n_{\rm obs}} + n_{\rm obs} \right)} - 1 \right]  &  {\rm if~S/N}_{\star} < {\rm S/N}_{\rm floor} \\
 & \\
\left( 1 - \frac{({\rm S/N}_{\star})^2}{({\rm S/N}_{\rm min})^2} \right)^{-1} \left[ 1 - \sqrt{ 1 - 2\left( 1 - \frac{({\rm S/N}_{\star})^2}{({\rm S/N}_{\rm min})^2} \right) } \right]  &  {\rm if~S/N}_{\star} = {\rm S/N}_{\rm floor} \\
\end{array}\right.
\end{equation}
In the case where S/N$_{\star}$~=~S/N$_{\rm floor}$, there is no
dependence on the number of light curve points because the photometric
error is not improved by adding more measurements. For a given set of
orbital parameters and planet radius, the fraction of the orbital
period during which the flux diminution exceeds this threshold is
proportional to the probability of observing that transit. By
appropriately averaging over all orbital phases and inclinations a
detection probability can be computed for each star.

The eclipse depth ($\delta$) for a uniformly illuminated stellar disk
is a function of the projected separation between the planet and star
($z$, which is in units of $R_{\star}$) and the ratio of their radii
($p$~$\equiv$~$R_{\rm p}/R_{\star}$)
\begin{equation} \label{eqn:dunif}
\delta = 
\left\{\begin{array}{lc}
0  &  z \geq 1+p \\
{1 \over \pi} \left[p^2 \kappa_0+\kappa_1-\sqrt{{4z^2-(1+z^2-p^2)^2\over 4}}\right] &  |1-p| < z < 1+p \\
 p^2 & z \le 1-p\\
1  &  z \le p-1\\
\end{array}\right.
\end{equation}
where $\kappa_0$~=~$\cos^{-1}[(p^2+z^2-1)/2pz]$ and
$\kappa_1$~=~$\cos^{-1}[(1-p^2+z^2)/2z]$ \citep{2002ApJ...580L.171M}.
The projected separation between the planet and star ($z$) is a
function of the orbital elements and phase
\begin{eqnarray}\label{eqn:z}
  z & = & \frac{a\sqrt{1-\sin^2i\cos^2\theta}}{R_{\star}}\left(\frac{1-e^2}{1-e\sin(\theta-\omega)}\right)
\end{eqnarray}
where $e$ is the eccentricity, $\omega$ is the argument of periastron,
$\theta$ is the orbital phase, and $i$ is the inclination
\citep{1983Ap&SS..92..203G}. Since we assumed circular orbits, the
rightmost term that depends on eccentricity becomes unity.

For a given inclination, the probability of observing a transit is the
fraction of the orbital period during which $\delta$~$>$~$\delta_{\rm
  min}$.  It is impossible to compute this directly because, by
inspection of Equations~\ref{eqn:dunif} and \ref{eqn:z}, the relevant
equations are transcendental; therefore, we computed this numerically.
First, we reduced the problem to three independent variables in
dimensionless form: $\delta_{\rm min}$, $R_{\star}/a$, and $R_{\rm
  p}/R_{\star}$.  For a given inclination, these variables uniquely
specify the range of orbital phase over which
$\delta$~$>$~$\delta_{\rm min}$, and we refer to this interval of
phase as $\Delta\theta_{\rm obs}$.  To compute the probability of
observing a transit ($P_{\rm obs}$), we numerically integrated these
phase intervals as a function of inclination, weighting by the
inclination probability distribution
\begin{eqnarray} \label{eqn:pobs}
  P_{\rm obs} & = & \int_0^{\pi/2} \frac{\Delta\theta_{\rm obs}(i)}{2\pi}~\cos{i} ~\,di
\end{eqnarray}
Because this is computationally intensive, we created a look-up cube
for the value of $P_{\rm obs}$ as a function of the three independent
variables ($\delta_{\rm min}$, $R_{\star}/a$, and $R_{\rm
  p}/R_{\star}$) and trilinearly interpolated this cube to assign
probabilities to the stars in our simulation. Since $P_{\rm obs}$
accounts for both the transit viewing angle probability
($R_{\star}/a$) and chance of catching the transit at some moment in
time ($\approx R_{\star}/\pi a$), it scales roughly as
$(R_{\star}/a)^2$. Of course, $P_{\rm obs}$ can be significantly
smaller than this if the minimum detectable depth ($\delta_{\rm min}$)
only allows a fraction of the transit to be detectable.

Finally, to determine the probability ($P_{\rm det}$) that a transit
would actually be detected in the survey, we used the binomial distribution
\begin{eqnarray} \label{eqn:pdet}
P_{\rm det} &=& \frac{n!}{n_{\rm obs}!(n-n_{\rm obs})!}P_{\rm obs}^{n_{\rm obs}}(1-P_{\rm obs})^{ n-n_{\rm obs}}
\end{eqnarray}
where $n$ is the total number of light curve points and $n_{\rm obs}$
is the number of points in which the transit was observed. For the
sparsely sampled light curves that we considered, we assumed that
transits are only ever observed once by the survey ($n_{\rm obs}$ =
1), and thus $P_{\rm det} \approx nP_{\rm obs}$ since $P_{\rm obs}$ is
small. The assumption that $n_{\rm obs}$ = 1 is valid because PS1 will
spend a total of only 6--12~minutes per bandpass over the 3-year
3$\pi$~Survey observing each star, which is a small fraction of the
typical orbital period (median 3.8~days). To calculate the survey
yield of detected transits, we simply summed the values of $P_{\rm
  det}$ determined for all the stars with assigned planets. The
results are given in Table~\ref{tbl:sim-res}, where we have used a
threshold of S/N~$>$~7 (see Section~\ref{sec:montcarl}). We found that
about 16 transits are detectable in the PS1 3$\pi$~Survey, with a
slight preference to be detected in $g$, $r$, and $y$ bands as fewer
nearby M~dwarfs are saturated in these bands compared to $i$ and $z$.
We have conservatively treated light curves from different bandpasses
separately (n.b., observations in different bandpasses are \emph{not}
observed simultaneously in PS1, unlike SDSS). However, it may be
possible to combine data from different bandpasses to enable more
robust rejection of false positives.

\subsection{Monte Carlo Simulation\label{sec:montcarl}}

While the semi-analytic approach allows a direct computation of the
transit yield, its limitations motivated us to perform a full Monte
Carlo simulation of survey light curves. One limitation is that we
implicitly assumed that the planet contributes no flux to the light
curve in the semi-analytic approach. Thus, without adding more
parameters it cannot be used to predict the survey yield of eclipsing
binaries, where the flux diminution during transit is reduced by the
light from a companion. In the semi-analytic approach we also could
not assess potential sources of false positives, the effects of
stellar limb-darkening, or how often systems are observed in eclipse
more than once.

In our Monte Carlo simulations, light curves were randomly drawn for
every survey star. The flux of each star was determined from its
magnitude in the PS1 bandpasses and the zero points derived in
Appendix~\ref{app:ps1}. Poisson noise was added to the light curves,
accounting for the flux diminution due to any eclipses. Eclipse light
curves were generated using the publicly available routines of
\citet{2002ApJ...580L.171M},\footnote{\url{\texttt{http://www.astro.washington.edu/users/agol/transit.tar.gz}}}
assuming either a uniformly illuminated stellar disk or a quadratic
limb-darkening law. We drew random orbital inclinations, and planet
and binary parameters were drawn from the distributions described in
Sections~\ref{sec:sim-comp} and \ref{sec:sim-plnt}. Limb-darkening
coefficients for the PS1 bandpasses were bilinearly interpolated from
the models of \citet{2004A&A...428.1001C} for the SDSS filters, which
are tabulated in \Teff\ and \logg. As shown in Appendix~\ref{app:ps1}
(Figure~\ref{fig:ps1-filt}), the SDSS and PS1 $griz$ bandpasses are
similar, and for the PS1 $y$-band we used the SDSS $z$-band
limb-darkening coefficients.

In order to assess eclipse false positives due to star spots, we added
sinusoidal variability to a fraction of the light curves. Since M
dwarfs are the primary source of detected eclipsing systems and can
potentially have large amplitude variability due to spots, we only
added simulated variability to M~dwarfs. We used the best available
information on such periodic variability from
\citet{2006MNRAS.367..407R}, who found a median variability amplitude
of 0.014~mag (0.005--0.02~mag observed range) in their $G$, $R$, and
$I$ bands for 21$\pm$11\% of M~dwarfs. In Appendix~\ref{app:alg}, we
discuss the issue of false positives in more detail, considering other
cases beyond our simple M~dwarf spot model.

\begin{figure} 
\centerline{\includegraphics[width=6.5in,angle=0]{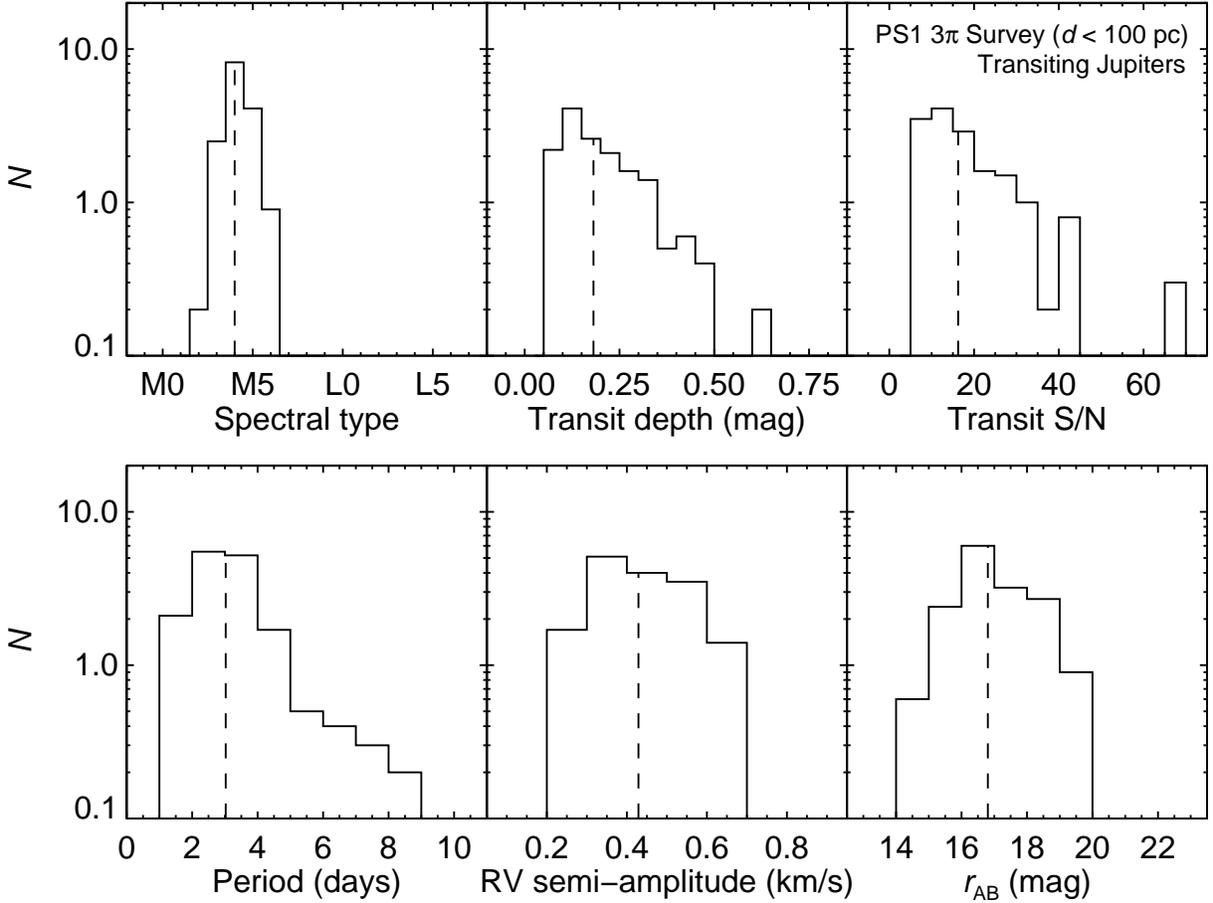}}

\caption{ \normalsize Properties of transiting Jupiter systems
  detected in our Monte Carlo simulations of the PS1 3$\pi$ Survey,
  considering only stars within 100~pc. The dashed lines indicate the
  medians of the distributions. A total of 13 transiting systems are
  detected around mid-M dwarfs (M3--M5) exclusively, with large
  typical flux diminutions of 0.15~mag (S/N$_{\rm transit}$ = 15). The
  typical orbital period of 3~days for a 1~\Mjup\ planet results in a
  radial velocity semi-amplitude of 0.4~\kms. (Note that histograms
  have noninteger values because the simulation was oversampled by a
  factor of 10.) \label{fig:sim-plnt}}

\end{figure}

\begin{figure} 
\centerline{\includegraphics[width=6.5in,angle=0]{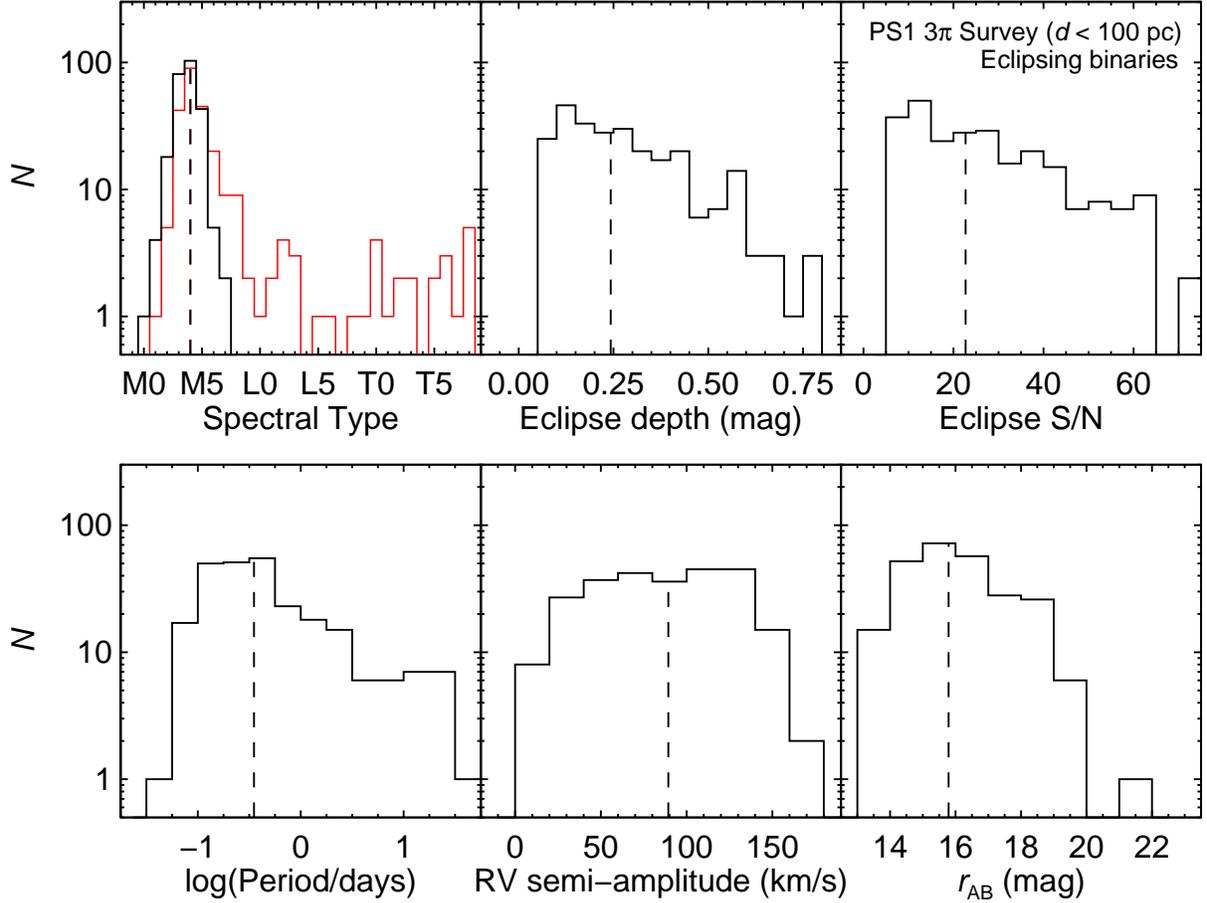}}

\caption{ \normalsize Properties of eclipsing binary systems detected
  in our Monte Carlo simulations of the PS1 3$\pi$~Survey, considering
  only stars within 100~pc. The dashed lines indicate the medians of
  the distributions. A total of 260 systems are detected, and the
  spectral types of the primaries are predominantly mid-M dwarfs
  (M3--M5, black line). The secondary components (red line) peak at a
  similar spectral type, but they extend to much later spectral
  types. Thus, $\sim$10 late-M and $\sim$10 L and T dwarfs are
  expected to be discovered in eclipsing binary systems by PS1,
  enabling the first direct measurements of the radii of field brown
  dwarfs.  Typical detected eclipse depths are large (0.25~mag,
  S/N$_{\rm transit}$ = 20), and the integrated magnitudes are
  sufficiently bright ($r \approx$ 15.8~mag) to measure precision
  radial velocities in the optical.  The typical orbital period of
  0.4~days results a large typical semi-amplitude of
  80~\kms. \label{fig:sim-comp}}

\end{figure}

We ran separate simulations for planets and binary companions (i.e.,
we did not consider transiting planets in double star systems) and we
used the ``S/N model'' (algorithm 3 in Appendix~\ref{app:alg}) in a
blind search of the simulated light curves for eclipse events. The
results are shown in Table~\ref{tbl:sim-res}, where we have again
treated bandpasses separately, as in our semi-analytic approach. We
adopted a threshold of S/N~$>$~7 in order to make the number of false
positives vanishingly small.\footnote{Note that if our choice of
  S/N~$>$~7 is an unrealistically optimistic threshold, this does not
  sensitively impact the planet or eclipsing binary yield.
  Figures~\ref{fig:sim-plnt} and \ref{fig:sim-comp} show that the
  median S/N of detected transits and eclipses are about 15 and 20,
  respectively.}  Summing the detections in individual bandpasses, the
survey yield for planets is in good agreement with the results from
the semi-analytic approach (both find 16 transits). We also found that
limb-darkened eclipse curves gave the same numbers to within
$\sim$10\% as uniform disk illumination, and thus in
Table~\ref{tbl:sim-res} we only quote the uniform disk results as
these are directly comparable to the semi-analytic results.

For the PS1 3$\pi$~Survey, the Monte Carlo simulations predict that
about 16 transits for 13 Jupiters orbiting M~dwarfs are detectable, so
about three transits are detected in two different bandpasses over the
course of the survey. However, the vast majority of detected flux dips
are predicted to be due to eclipsing binaries: 570 eclipses for 260
eclipsing binaries. About 40\% of these binaries are detected in only
one bandpass, about 25\% are detected in two, and about 15\%/10\%/10\%
are detected in 3/4/5 bands. Because different bandpasses are not
observed simultaneously, such multi-band detections cannot be used to
measure the color in and out of eclipse, which could help distinguish
planets from stellar companions.

Figure~\ref{fig:sim-plnt} shows that the host stars of detected
transiting Jupiters are overwhelmingly mid-M~dwarfs (M3--M5), which is
not surprising as these are the most common stars in a volume-limited
sample \citep{2004AJ....128..463R}. Figure~\ref{fig:sim-comp} shows
that the primary stars of detected eclipsing binaries are also
predominantly mid-M~dwarfs, and the secondaries range from mid-M to L
and T dwarfs. Our simulations predict $\sim$10 eclipsing binaries with
a substellar secondary, and these would be important discoveries as
there are currently no known field brown dwarfs in eclipsing binaries
and therefore no direct measurements of the radii of brown dwarfs.

Different bandpasses contribute different numbers of eclipsing
binaries by a factor of $\sim$2. For example, the $g$ band contributes
about twice as many eclipses as the $i$ or $z$ bands. This is a
natural consequence of the fact that M~dwarfs have red optical colors
and the detection volume of mid-M~dwarfs within 100~pc is most
strongly limited by saturation. Thus, the effective volume of $g$ band
is $\sim$2 larger than $i$ or $z$ bands because saturation affects $g$
band less (see Figure~\ref{fig:survey-limits} in
Appendix~\ref{app:ps1}).


\section{Discussion\label{sec:disc}}

Our simulations have shown that a large population of eclipsing
binaries, in which mid-M~dwarfs are the primary star, are likely to be
detected by multi-epoch all-sky surveys such as the PS1 3$\pi$~Survey.
Given reasonable assumptions for the planet fraction, we also predict
that a substantial number of transiting Jupiters may be detectable,
also around mid-M host stars. We now discuss how our results are
impacted by the input assumptions to our simulations, as well as the
prospects for confirming candidate eclipsing systems using follow-up
observations.

\subsection{Impact of Assumptions on Survey Yield \label{sec:assump}}

\begin{itemize}

\item \emph{M~dwarf planet frequency.}  The transit yield depends
  directly on the assumed planet frequency.  Currently, no M~dwarfs
  are known to host a transiting Jupiter-mass planet, so this fraction
  is very uncertain. Several M~dwarfs are known to host planets, about
  half of which are less massive than Neptune
  \citep{2009ApJ...690..743B}.  Such a preference of lower mass stars
  to form lower mass planets was predicted by
  \citet{2004ApJ...612L..73L}, and subsequent models
  \citep{2005ApJ...626.1045I, 2008ApJ...673..502K} now also reproduce
  the observational result of \citet{2007ApJ...670..833J} that
  Jupiter-mass planets are less common around lower mass
  stars.\footnote{Such models typically rely on a correlation between
    the stellar mass and protoplanetary disk mass ($M_{\rm disk}
    \propto M_{\star}$).  However, observations of circumstellar disks
    show no such correlation \citep[e.g.,][]{2005ApJ...631.1134A}, and
    models that do not include this correlation find that planets are
    \emph{more} common around lower mass stars
    \citep{2006A&A...458..661K}.}  Some fraction of M~dwarfs do form
  Jupiter-mass planets, and there are currently several examples of
  such systems discovered by RV and microlensing surveys
  \citep{1998A&A...338L..67D, 1998ApJ...505L.147M,
    2004ApJ...617..580B, 2007ApJ...670..833J, 2009ApJ...690..743B,
    2009ApJ...695..970D}. We have simply scaled the fraction of
  short-period Jupiters from solar-type host stars (1.1\%) to M~dwarfs
  (0.47\%) using the relative Jupiter-mass planet fraction found by
  \citet{2007ApJ...670..833J}. However, due to the uncertainty in the
  different formation and migration histories of planets orbiting
  solar-type stars and M~dwarfs, this assumption may not be correct.
  Ultimately, our findings suggest that the PS1 3$\pi$~Survey will be
  capable of addressing this question since it would find $\gtrsim$~10
  transiting Jupiters if this simple scaling is correct and
  $\gtrsim$~20 if the population were identical to solar-type
  stars. Thus, even a null result would place meaningful upper limits
  on the population of short-period Jupiters around M~dwarfs.

\item \emph{M~dwarf planet orbital parameters.}  Our assumptions about
  the orbital parameters of short-period Jupiters around M~dwarfs also
  affect the transit yield.  We have assumed a period distribution
  identical to that of all observed short-period planets, which is
  dominated by the planetary systems of solar-type stars.  The short
  period ``pile-up'' in the planet period distribution is commonly
  attributed to the halting of planetary migration because of an inner
  edge to the circumstellar disk \citep[e.g.,][]{2002ApJ...574L..87K}.
  Thus, M~dwarfs may have a systematically different period at which
  planets terminate their migration.  Given their lower energy output,
  the inner disk edge for M~dwarfs should be closer to the star, which
  would result in a shorter pile-up period and an increase in the
  number of detected transits. For example, if the inner disk edge
  occurs at a fixed temperature, where the gas becomes ionized
  \citep{2000prpl.conf..589S} and/or dust sublimates
  \citep{2000ApJ...543..486S}, its distance from the star will scale
  as $a~\propto~\sqrt{L_{\star}}$.  During pre--main-sequence
  evolution $L_{\star}~\propto~M_{\star}^2$
  \citep{2000A&A...358..593S}, so the location of the inner disk edge
  is expected to scale as $a~\propto~M_{\star}$ and the orbital period
  as $P~\propto~\sqrt{a^3/M_{\star}}~\propto~M_{\star}$. Short-period
  planets around solar-type stars ($P$~$<$~11.5~days,
  0.7~$<$~$M_{\star}$~$<$~1.3~\Msun) in the Catalog of Nearby
  Exoplanets \citep{2006ApJ...646..505B} have a median orbital period
  of 3.9~days, thus such planets around mid-M~dwarfs (0.3~\Msun) would
  be expected to have a median orbital period of 1.3~days. This would
  drastically change the predicted number of detectable transiting
  planets in our simulations. We ran a simulation in which we assumed
  orbital periods scaled with mass for all M~dwarfs, and the number of
  predicted transit detections increased by a factor of $\sim$4.

  We have also assumed an eccentricity of zero for planets, as well as
  eclipsing binaries, which is expected from tidal circularization of
  such short-period orbits
  \citep[e.g.,][]{1977A&A....57..383Z}. However, if the orbits
  possessed modest eccentricities such as those observed for longer
  period planets ($P$~$>$~10~days, median eccentricity of 0.3),
  \citet{2008ApJ...679.1566B} has shown that this would result in a
  25\% increase in transit probability.

\item \emph{Mass function.}  Both the eclipsing binary and transit
  yield depend strongly on the functional form and overall
  normalization of the mass function, as these determine the
  proportion of M~dwarfs and total number of stars, respectively, in
  our simulations. As described in Section~\ref{sec:sim-star}, we have
  adopted the log-normal mass function of \citet{2008AJ....136.1778C},
  and they cite a range of possible values for the best fit mass
  function parameters: 0.08--0.12~$\log(M/\Msun)^{-1}$~pc$^{-3}$ for
  the normalization ($A$), 0.20--0.50~\Msun\ for the peak ($M_c$), and
  0.22--0.37 for the width ($\sigma_{\log{M}}$). If we vary the
  parameters within these limits the number of objects within 100~pc
  varies from 2.7$\times$10$^5$ to 5.0$\times$10$^5$ (i.e., about
  $\pm$30\%).  Thus, the yields from our simulations are
  correspondingly uncertain by $\pm$30\% from the normalization of the
  mass function alone.

  We also tried the power-law mass function derived by
  \citet{1999ApJ...521..613R}, which predicts 4.1$\times$10$^5$
  low-mass stars and brown dwarfs (0.01~$<$~$M/\Msun$~$<$~1.0) within
  100~pc.  This is very similar to the total number derived from the
  log-normal mass function because, as \citet{2008AJ....136.1778C}
  pointed out, the data used to derive both functions are in good
  agreement even though the analytic expressions adopted by the
  authors differ. These different analytic expressions result in
  objects being apportioned differently over the mass range we
  consider (0.01~$<$~$M$~$<$~1.0~\Msun). The essentially flat (in
  $\log{M}$) power-law mass function of \citet{1999ApJ...521..613R}
  results in roughly an equal number of objects in the range
  0.01--0.1~\Msun\ as in the range 0.1--1.0~\Msun, whereas the
  log-normal mass function, which peaks at 0.29~\Msun, results in more
  than 10$\times$ the number of objects in the upper range compared to
  the lower range.  Thus, the choice of mass function can change the
  number of low-mass stars by a factor of $\sim$2 and the number of
  brown dwarfs by a factor of $\sim$10.\footnote{In fact, we chose to
    adopt the log-normal mass function because it is the most
    consistent with the best available data for M~dwarfs
    \citep{2008AJ....136.1778C}, and it is also consistent with the
    number density of T~dwarfs in the solar neighborhood
    \citep{2008ApJ...676.1281M, 2008MNRAS.390..304P}, whereas a
    power-law mass function flat in $\log{M}$ predicts significantly
    more T dwarfs than are observed.}  Correspondingly, we found that
  the eclipsing binary yield was reduced by about 50\% (as expected),
  while the reduction in the transit yield was smaller (about 30\%)
  because the increase in later type M~dwarfs somewhat compensated for
  the reduction in mid-M~dwarfs.  In addition, the factor of $\sim$10
  increase in the numbers of brown dwarfs resulted in $\sim$3
  predicted detections of substellar eclipsing binaries not seen when
  using the log-normal mass function.

\item \emph{Binary parameters.}  The eclipsing binary yield also
  depends on the binary properties we have adopted for M~dwarfs, which
  is ultimately based on the period distribution of
  \citet[][DM91]{1991A&A...248..485D} and the mass ratio distribution
  of \citet[][FM92]{1992ApJ...396..178F}. The FM92 mass ratio
  distribution has not been corrected for observation bias or
  completeness and is dominated by wide, long-period systems.  A
  recent compilation of known low-mass eclipsing binaries
  \citep{2007ApJ...660..732L} shows the clear preference of such
  systems toward mass ratios of unity, and in fact, this phenomenon of
  short-period ``twins'' is expected from binary formation theory
  \citep{1983AcA....33..223A, 2000MNRAS.314...33B}.  Having more twins
  would mean that more M~dwarf eclipsing binaries would be easier to
  detect (as the larger companion radii would lead to deeper
  eclipses), but that there should be fewer (if any) brown dwarfs
  found in eclipsing systems with M~dwarfs.

  The assumed binary period distribution can also affect the predicted
  yield significantly, since the number of eclipsing binaries strongly
  depends on its short-period tail. Therefore, we also tried adopting
  a compressed DM91 distribution for M~dwarfs, similar to the approach
  of \citet{2006AJ....132..663B} for very low-mass stars, who based
  their distribution on the models of \citet{2004ApJ...600..769F} that
  predict shorter period orbits for lower mass binaries. Using a DM91
  period distribution shifted and compressed by 0.5 in $\log{P}$
  (i.e., assuming typical masses for the FM92 and DM91 samples of
  $\sim$0.3~\Msun\ and $\sim$1.0~\Msun), we found essentially the same
  eclipsing binary yield. Although the tails of the distribution were
  reduced, the shift to shorter periods roughly compensated, yielding
  the same number of short-period M~dwarf binaries.

\item \emph{Substellar binaries.}  Although our simulations do not
  predict any eclipsing binaries composed solely of brown dwarfs, the
  scarcity of data for brown dwarf spectroscopic binaries renders the
  orbital period distribution, particularly at short periods, very
  uncertain.  In fact, combined with the uncertainty in the substellar
  portion of the mass function, some brown dwarf eclipsing binaries
  may indeed be detected by the PS1 3$\pi$~Survey. Because of the flat
  mass--radius relation for brown dwarfs, substellar eclipsing
  binaries should be among the easiest eclipsing systems to detect:
  equal radii imply eclipse depths of 50--100\% (depending on the flux
  ratio) over a wide range of mass ratios.

\item \emph{Survey cadence.}  Our assumption of uniform phase coverage
  may impact the survey detection sensitivity.  In particular, for
  each bandpass the PS1 3$\pi$~Survey as planned will obtain images in
  pairs separated by 30--60~minutes for the purpose of asteroid
  detection.  In our simulations, we have treated these two epochs as
  independent because this interval is comparable to or larger than
  the expected eclipse durations.  For systems detected in our
  simulations, median eclipse durations were 45 minutes for transiting
  planets and 30 minutes for eclipsing binaries.  Thus, it is unlikely
  (but not impossible) that an eclipse event could be detected in both
  images of a 3$\pi$~Survey observation pair, which would enable more
  robust detections of candidate eclipsing systems.  We have
  conservatively chosen not to model this effect, although it may play
  an important role in both eclipse detection and false positive
  rejection in actual survey data.

\item \emph{Seeing/saturation.} We have assumed typical seeing of
  0\farcs8 for PS1, located on Haleakala, on the island of Maui
  (Appendix~\ref{app:ps1}). Current on-sky assessments of the image
  quality indicates a more realistic estimate may be 1\farcs0.  The
  seeing has a significant impact on the saturation limits, and such a
  change would make these less restrictive by 0.4~mag.\footnote{Seeing
    of 1\farcs0 also changes the S/N~=~5 detection limits, by 0.2~mag,
    but this does not significantly impact our results as M~dwarfs
    within 100~pc are well above the detection threshold.} This would
  make more of the earlier type and/or nearer M~dwarfs accessible to
  the 3$\pi$~Survey, increasing the number of detectable transiting
  planets and eclipsing binaries.  For example, Table~\ref{tbl:100pc}
  shows that 22\% of M~dwarfs within 100~pc are saturated assuming
  0\farcs8 seeing, and this would change to 15\% for 1\farcs0 seeing.
  Thus, more M~dwarfs would be unsaturated in more bandpasses.

\end{itemize}

\subsection{Follow-up Observations \label{sec:follow}}

Eclipse events are only expected to be observed once, or possibly
twice, per star over the entire course of the PS1 3$\pi$ survey. This
has two consequences: (1) the eclipse signatures of transiting planets
and eclipsing binaries will be indistinguishable from each other in
PS1 data; and (2) detailed photometric and spectroscopic follow-up
observations will be required to characterize all systems.  Since
eclipsing binaries are predicted to outnumber transiting planets by
$\sim$20:1 in survey data, the initial step of distinguishing between
these two classes of objects, which have very different RV precision
requirements, should be done as efficiently as possible. Since all of
the objects we have considered will have parallaxes measured by PS1,
it may be possible to distinguish eclipsing binaries from transit
hosts since they will appear overluminous on the color--magnitude
diagram, especially if the binaries are preferentially twins (as
discussed in Section~\ref{sec:assump}).  However, we conservatively
assume that robust discrimination will only be possible with
photometric and/or spectroscopic follow-up observations.

Transiting planets may be distinguished from binaries either by their
much lower RV semi-amplitude (via spectroscopic follow-up) or by the
shape of the light curve and lack of a secondary eclipse (via
photometric follow-up). With unknown ephemerides, RV measurements from
at least two, possibly three, epochs would be needed to identify
eclipsing binaries. Assuming 20 pairs of RV measurements could be
obtained per night, at least 14 nights would be required to follow-up
all 273 candidates. This would be a substantial investment of
resources, even if the brightest targets could be done with smaller
aperture telescopes (e.g., CFHT/Espadons) instead of large apertures
(e.g., Keck/HIRES). In fact, this RV follow-up would be much more
powerful if it were obtained \emph{after} ephemerides were determined
from photometric follow-up. The RV observation epochs could then be
optimally timed to sample maxima in the RV curves, and these two RV
measurements along with a light curve would yield a full orbital
solution (assuming circular orbits).

Photometric follow-up would be most efficiently broken into two
campaigns: first, to determine rough ephemerides (transit times and
periods); second, to measure higher precision and cadence light curves
during transit.  The initial campaign would require regular monitoring
every eclipse half-interval (15 minutes) over at most two full periods
(0.4~days for binaries; 3~days for planets), in order to catch two
eclipses.  Assuming 5 minutes per observation (slewing and integration
time), this amounts to 6.4~hours per binary and 48 hours per planet,
for a total of 230 nights.  With ephemerides in hand, detailed light
curves can then be obtained.  In order to measure both the primary and
secondary eclipses of binaries, at least two such light curves are
required, so this second phase would cost $\approx$2~hours per object,
for a total of 55 nights. Thus, all the necessary photometric
follow-up could cost as much as 285 nights, essentially all the clear
nights in a year for a single telescope.  However, particularly for
the initial campaign of 230 nights, this cost could be distributed
among several small telescopes, possibly even in collaboration with
amateur astronomers -- a strategy employed with success by the XO
transit survey \citep{2006ApJ...648.1228M} and the
MicroFUN\footnote{\url{\texttt{http://www.astronomy.ohio-state.edu/$\sim$microfun/}}}
consortium.  This is possible because eclipsing systems discovered in
the 3$\pi$ Survey would be relatively bright: PS1 achieves a
S/N~$\gtrsim$~100 in 30~s with a 1.8-m telescope, so $\approx$30~cm
apertures should be able to achieve a photometric precision of
$\approx$2\% in just a few minutes.

To summarize the initial follow-up requirements, either an RV campaign
costing at least 14 nights on 8-m class telescopes or a photometric
campaign costing 285/$N$ nights on $N$ small-aperture
($\gtrsim$~30~cm) telescopes is needed to separate all the transiting
planets from the 20:1 more numerous eclipsing binaries. We suggest
that the photometric monitoring campaign be conducted first, so that
eclipsing binaries can be identified by the presence of a shallower
secondary eclipse, and also so that ephemerides can be determined for
both the transiting planets and eclipsing binaries. A subsequent RV
monitoring campaign can then focus only on the planets and a subset of
the most interesting eclipsing binaries, thus reducing the amount of
large-aperture telescope time needed.

For the eclipsing binaries that are predicted to be discovered by the
PS1 3$\pi$ Survey, the RV precision necessary for precise parameter
determination is attainable with current instrumentation. Our
simulations predicted a median binary orbital period of 0.4~days,
semimajor axis of 0.006~AU, primary mass of 0.2~\Msun, mass ratio of
0.7, and inclination of 86\degs, resulting in a typical RV
semi-amplitude ($K_1$) of 80~\kms.  This is very similar to the
eclipsing binary for which \citet{2008MNRAS.386..416B} recently
achieved $\sim$3~\kms\ precision using Keck/HIRES, even though their
target ($r$~=~18.3~mag) is $\approx$2.5~mag fainter than those
predicted to be found by PS1 (median $r$~=~15.8~mag;
Figure~\ref{fig:sim-comp}).

For transiting Jupiters the RV precision requirements are much more
demanding. Our simulations predicted a median planet orbital period of
3~days, semimajor axis of 0.02~AU, host star mass of 0.2~\Msun\ (and
thus a mass ratio of 0.005), and inclination of 89\degs, resulting in
a typical RV semi-amplitude of 0.4~\kms. While current planet searches
have used optical spectroscopy to achieve a precision of 5--10~\ms\
for M~dwarfs \citep[e.g.,][]{2004ApJ...617..580B,2005A&A...443L..15B},
such targets are $\gtrsim$~100$\times$ brighter than the M~dwarf host
stars from our simulations. For an M~dwarf of comparable brightness to
such PS1 candidates, \citet{2008ApJ...678L.125B} have recently
demonstrated a precision of 0.3~\kms\ using high-resolution
near-infrared spectroscopy. While this is the current state of the
art, much better precision will be needed to measure the RV signal of
transiting Jupiters around M~dwarfs. Future near-infrared
spectrographs aim to reach a precision of 1~\ms\ (i.e., comparable to
that of current optical spectrographs), which would enable the
detection of planets around M~dwarfs down to an Earth mass
\citep[e.g.,][]{2008SPIE.7014E..31J}.

\subsection{Extending the Volume-limited Sample  \label{sec:extend}}

\begin{figure} 
\centerline{\includegraphics[width=6.5in,angle=0]{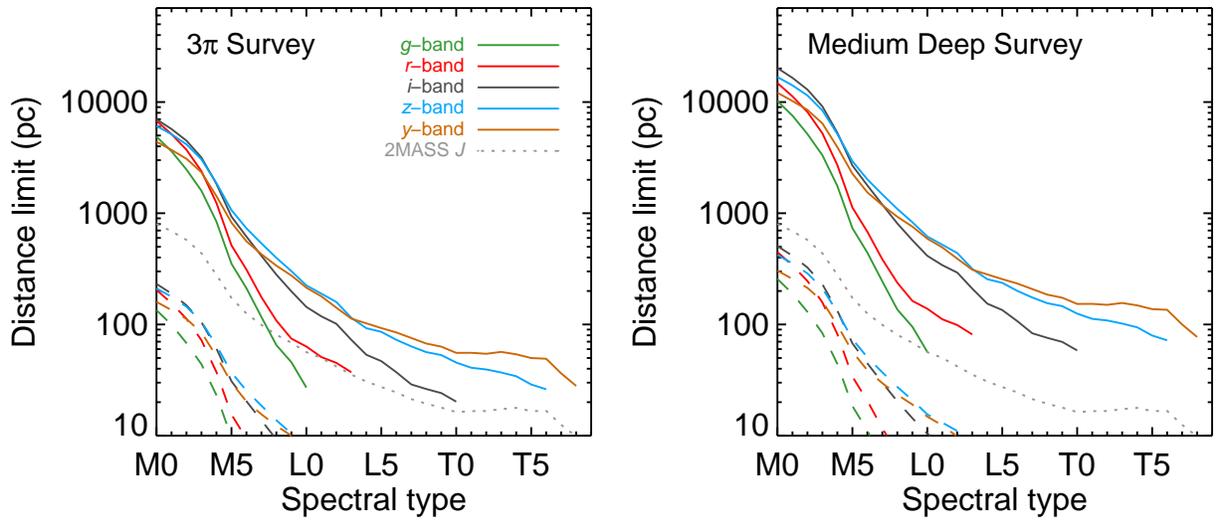}}

\caption{ \normalsize Distance at which objects of a given spectral
  type reach the survey detection limit (S/N~=~5) in a given bandpass
  (solid lines). For comparison, the dotted line shows the detection
  limits for 2MASS ($J$~$<$~15.8~mag). The dashed lines show the
  distances within which objects are expected to be saturated. The
  solid lines terminate at spectral types beyond which there is
  insufficient data to predict detection limits, as there is little
  published optical photometry or
  spectroscopy. \label{fig:survey-limits}}

\end{figure}

In the case of the PS1 3$\pi$~Survey, we have only considered a
volume-limited sample extending to 100~pc, the distance within which
parallaxes will be reliably measured. However, M~dwarfs remain
detectable at even greater distances (see
Figure~\ref{fig:survey-limits}), so one could considerably expand the
search volume and in turn the number of eclipse/transit detections.
Although objects in such an expanded sample will be more difficult to
follow-up due to their faintness, it may be the only way to discover
especially rare objects such as substellar eclipsing binaries.

Optical and/or near-infrared colors can reliably discriminate between
nearby M type dwarfs and background giants
\citep[e.g.,][]{1988PASP..100.1134B, 2003AJ....126.2421C}. Thus,
combining the PS1 optical bandpasses with photometry from 2MASS may
enable a robust color-selected sample of M~dwarfs out to the detection
limit of the shallower survey, which in this case is 2MASS
(Figure~\ref{fig:survey-limits}). The 2MASS point source catalog is
complete for mid-M~dwarfs to a distance of $\approx$200~pc (130~pc for
M6, 280~pc for M4). Expanding the volume-limited sample to this
distance would increase the survey volume by a factor of $\approx$8.

The PS1 3$\pi$~Survey will of course measure proper motions for many
more stars than it will measure parallaxes. Assuming an astrometric
precision of 10~mas per epoch and an operational lifetime of 3~years,
PS1 will be able to measure proper motions of
$\gtrsim$~0\farcs03~yr$^{-1}$ (S/N~$>$~10).  M~dwarfs in the
volume-limited sample of \citet{1995AJ....110.1838R} have a median
tangential velocity of about 40~\kms, which translates to
0\farcs03~yr$^{-1}$ at a distance of 300~pc. The PS1 3$\pi$~Survey
photometric detection limits for mid-M~dwarfs are well beyond 300~pc
(e.g., M6 stars can be detected to 770~pc). Thus, a proper motion
selected sample would contain most of the M~dwarfs within 300~pc,
which is about $\gtrsim$~20$\times$ more M~dwarfs than in the parallax
selected sample. However, eclipsing systems discovered in such an
extended sample will be $\sim$10$\times$ fainter than objects
discovered within 100~pc, which will make spectroscopic follow-up more
challenging.


\section{Conclusions\label{sec:conc}}

We have developed detailed simulations to predict the yield of
eclipsing systems in sparsely sampled multi-epoch all-sky surveys, and
we have applied these to the Pan-STARRS-1 (PS1) 3$\pi$~Survey. We
created a model of the solar neighborhood that includes low-mass
stars, brown dwarfs, and a realistic population of binary
companions. We devised a method to compute the number of transiting
planets in this sample semi-analytically, in addition to simulating
actual survey light curves using Monte Carlo methods. In
Appendix~\ref{app:alg}, we have investigated various algorithms for
searching sparsely sampled light curves for eclipse events and
carefully considered the vulnerability of these techniques to the most
likely source of astrophysical false positives, star spots.

Because upcoming multi-epoch surveys such as PS1 will achieve an
astrometric precision of 10~mas per epoch, they will measure
parallaxes for the hundreds of thousands of stars within 100~pc. Such
an unprecedented sample of nearby stars will provide the means for an
innovative approach to eclipse hunting by focusing only on the
closest, best characterized stars. In our simulations, we have only
considered this 100-pc sample, although it may be possible to expand
the search volume -- and thus the predicted survey yields -- by an
order of magnitude or more by using color selection and/or the proper
motions also measured by these surveys. Restricting an eclipse search
to a volume-limited sample naturally reduces the exotic astrophysical
false positives that have afflicted previous transit surveys. However,
the gain in areal coverage comes at the price of sparse light curve
sampling, resulting in single-epoch eclipse detections. Thus, a low
rate of photometric outliers in survey data is essential, and
extensive follow-up observations will be needed to characterize
candidate eclipsing systems.

A volume-limited sample is dominated by the most abundant stars:
M~dwarfs. Assuming an intrinsic frequency of 0.47\% for Jupiter-like
planets in short-period orbits ($P$~$<$~11.5~days) around M~dwarfs,
our simulations predict that about 13 transiting planets would be
detectable in the PS1 3$\pi$~Survey. Under the same assumptions, an RV
survey would need to monitor $\sim$30,000 M dwarfs to discover an
equivalent number of transiting Jupiters around M dwarfs.  Although
the host stars of the transiting Jupiters discovered by PS1 would be
$\sim$100$\times$ fainter than the M~dwarf hosts of known extrasolar
planets, they would be sufficiently bright to permit spectroscopic
follow-up with currently available near-infrared spectrographs. The
properties of planets around M~dwarfs are poorly constrained by
current observations, and the expected planet yield is correspondingly
uncertain. For example, if we assume a simple linear scaling with host
star mass of the inner edge of protoplanetary disks, which is thought
to define the terminus of planet migration, the orbital periods of
close-in M~dwarf planets would be $\sim$3$\times$ shorter than their
solar-type counterparts, resulting in a factor of $\sim$4 increase in
the transit yield from our simulations.

Our simulations also predict that hundreds of low-mass eclipsing
binaries with individual masses $<$~0.5~\Msun\ are potentially
detectable in the PS1 3$\pi$~Survey 100-pc sample. This would be a
major step forward in testing models of low-mass stars, which
currently rely on mass and radius measurements from only seven such
systems. In fact, only a few of these systems have parameters
determined to high precision because of their apparent faintness,
whereas eclipsing binaries discovered in a volume-limited sample by
PS1 would naturally include some of the brightest eclipsing binaries in the
sky. The number of detectable substellar eclipsing binaries is
uncertain (ranging from none to $\sim$3 for the 3$\pi$~Survey) due to
poor observational constraints on substellar regime of the initial
mass function and on the properties of brown dwarf binaries at such
small orbital separations. The mass ratio distribution for tight
M~dwarf binaries is nearly as uncertain, but using the best available
data our simulations predict that dozens of brown dwarfs are
potentially detectable in eclipsing binaries with M~dwarfs. Such
systems would enable radius measurements of field brown dwarfs for the
first time, providing sorely needed tests of substellar evolutionary
models.

This is the first detailed analysis of the expected yield of eclipsing
systems from Pan-STARRS, revealing its potential to discover systems
that have largely eluded previous surveys. The extensive set of tools
we have developed for our simulations can be readily applied to other
sparsely sampled, multi-epoch all-sky surveys planned for the coming
decade. Surveys such as LSST are anticipated to far exceed the
revolutionary surveys of the present by imaging the whole sky deeper
and more often, opening the door to intrinsically fainter and rarer
eclipsing systems.


\acknowledgments

We have benefited from conversations with Eugene Magnier about the PS1
surveys, J. Patrick Henry and Thierry Forveille about statistics, Eric
Hilton about M~dwarf variability and flares, and Adam Kraus about
low-mass binaries. We are grateful to John Tonry for providing
information about the PS1 photometric system.
Our research has employed NASA's Astrophysical Data System; the SIMBAD
database operated at CDS, Strasbourg, France; the SpeX Prism Spectral
Libraries, maintained by Adam Burgasser at
\texttt{http://www.browndwarfs.org/spexprism}; Sandy Leggett's L and T
dwarf archive at \texttt{http://www.jach.hawaii.edu/~skl/LTdata.html};
and the M, L, and T~dwarf compendium housed at
\texttt{DwarfArchives.org} and maintained by Chris Gelino, Davy
Kirkpatrick, and Adam Burgasser \citep{2003IAUS..211..189K,
  2004AAS...205.1113G}.
TJD and MCL acknowledge support for this work from NSF grant
AST-0507833, and MCL acknowledges support from an Alfred P. Sloan
Research Fellowship.


\clearpage
\appendix

\section{The Pan-STARRS-1 3$\pi$ and Medium Deep
  Surveys\label{app:ps1}}

In the following, we describe the attributes of the PS1 photometric
system and the 3$\pi$ and Medium Deep surveys. PS1 is a 1.8-m
telescope located on Haleakala on the island of Maui, and has an
effective collecting area of 1.73~m$^{-2}$, after accounting for
obscuration and diffraction losses. It is equipped with the Gigapixel
Camera \#1 (GPC1), comprising 60 orthogonal transfer array (OTA)
chips, each 4k$\times$4k pixels, sampling a circular field-of-view
3\degs\ in diameter.  GPC1 has a nominal pixel scale of
0\farcs26~pixel$^{-1}$, read noise ($\sigma_r$) of
5~$e^-$~pixel$^{-1}$, and gain of 1~$e^-$~DN$^{-1}$
\citep{2008SPIE.7014E..12O}. We have assumed that the best possible
photometric precision for PS1 is 0.01~mag, i.e., a systematic noise
floor of (S/N)$_{\rm floor}$~=~100 in the photometry.

The PS1 photometric system is largely patterned after the SDSS
photometric system \citep{1996AJ....111.1748F}, but the PS1 detectors
are much more red-sensitive so low-mass stars and brown dwarfs will
have somewhat different colors in the PS1 photometric system than in
SDSS. The total predicted system throughput in each of the five PS1
bandpasses ($grizy$) is shown in Figure~\ref{fig:ps1-filt} along with
the SDSS bandpasses. Compared to SDSS, PS1 has significantly increased
sensitivity at $z$-band, along with an additional bandpass ($y$)
centered at 0.98~\micron. We have computed the AB system zero points
of these bandpasses, as well as the expected sky flux in each filter.
These quantities are listed in Table~\ref{tbl:ps1-filt}.

For our simulations, we drew on the extensive body of optical
photometry from SDSS for low-mass stars and brown dwarfs
\citep{2002astro.ph..4065H, 2004AJ....127.3553K, chiu05}. Therefore,
we have derived conversions from the SDSS to the PS1 photometric
systems as a function of spectral type by computing synthetic
photometry from publicly available spectra. We used the averaged
stellar spectra of G0--M6 dwarfs from \citet{1998PASP..110..863P},
which has essentially complete wavelength coverage at optical and
near-infrared wavelengths.  We used optical spectra retrieved from
\url{\texttt{dwarfarchives.org}} for K5--M9 dwarfs, most of which are
either unpublished or are from \citet{1991ApJS...77..417K},
\citet{1994AJ....108.1437H}, or \citet{1995AJ....109..797K}. We used
flux-calibrated spectra of M, L, and T dwarfs both from
S.~K.~Leggett's spectral
library,\footnote{\url{\texttt{http://www.jach.hawaii.edu/$\sim$skl/LTdata.html}}}
which comprise a variety of published spectra, and from the IRTF/SpeX
spectral library \citep[][J.~T.~Rayner et al. 2009,
submitted]{2005ApJ...623.1115C}.\footnote{\url{\texttt{http://irtfweb.ifa.hawaii.edu/$\sim$spex/IRTF\_Spectral\_Library/}}}

\begin{figure} 
\centerline{\includegraphics[width=3.5in,angle=0]{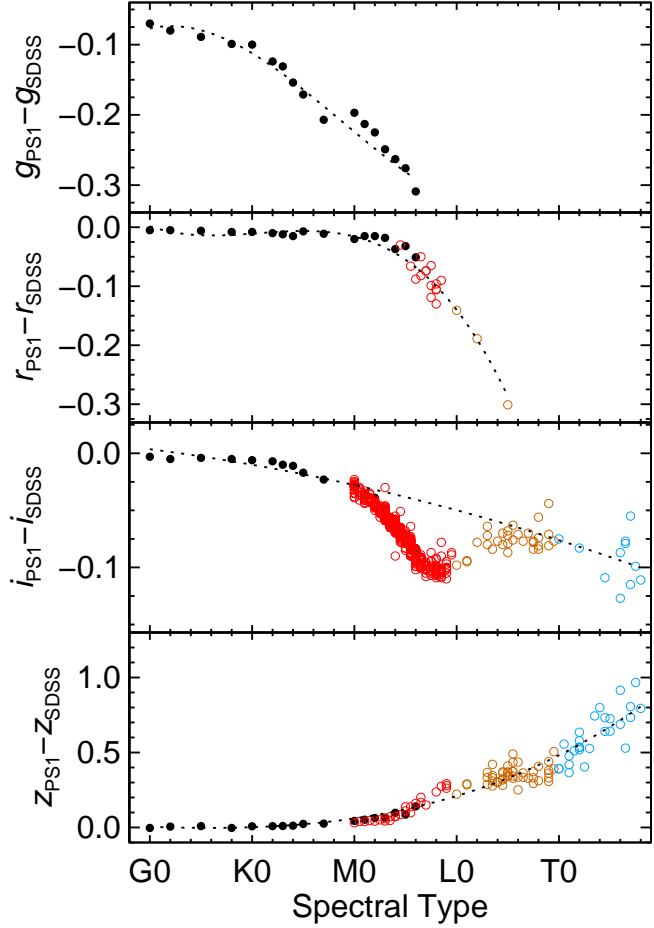}}

\caption{ \normalsize Conversions in magnitudes between the SDSS and
  PS1 photometric systems, derived from synthetic photometry as
  described in the text. Open circles indicate spectra of individual
  objects, with M, L, and T dwarfs show in red, brown, and blue,
  respectively.  Filled black circles indicate the average stellar
  spectra of \citet{1998PASP..110..863P}.  Polynomial fits to these
  data are shown by the dotted lines (coefficients are given in
  Table~\ref{tbl:phot-conv}). \label{fig:phot-conv}}

\end{figure}

Our derived SDSS-to-PS1 photometric conversions are shown in
Figure~\ref{fig:phot-conv}, along with polynomial fits as a function
of spectral type (coefficients given in Table~\ref{tbl:phot-conv}).
These conversions reflect how the flux ratio between an object of a
given spectral type and a zero-magnitude source differs from SDSS to
PS1. By definition, the AB zero-magnitude source is blue
($F_{\lambda}~\propto~\lambda^{-2}$), while M, L, and T dwarfs are red
at optical wavelengths. Thus, PS1 filters that are somewhat redder
than their SDSS counterparts (like $g$ and $r$) gain object flux
without as much change in the zero-magnitude flux, leading to brighter
PS1 magnitudes of up to 0.3~mag for the reddest objects. The PS1 $i$
band is very similar to SDSS, resulting in $\lesssim$~0.1~mag
differences in zero point.\footnote{We note that a simple polynomial
  fit to the $i$ band synthetic photometry does not accurately
  reproduce details in the $i$ band photometric conversion as a
  function of spectral type.  However, the discrepancies between our
  fit and the actual photometry are $\lesssim$~0.06~mag, and thus have
  an insignificant impact on our results.} The biggest photometric
conversion factor is for the $z$ band, which is understandable since
for SDSS this is a longpass filter but for PS1 it has well defined
cut-on and cut-off wavelengths.

The Pan-STARRS-1 Telescope will be used to conduct several surveys,
and the largest two of these are the 3$\pi$~Survey and the Medium Deep
Survey, nominally using 56\% and 25\% of the available observing time,
respectively. At the time of writing, the exact survey strategies to
be employed are somewhat uncertain, pending on-sky tests of PS1
performance. For our simulations, we have assumed the surveys will be
conducted as conceived in the Design Reference Mission document. The
cadence employed by the 3$\pi$~Survey is planned to be optimized for
parallax measurements of nearby low-mass objects, with redder bands
sampling epochs nearer the maximum parallax factor. Thus, observations
in each of the five bandpasses will be spread out over each observing
season. Also, each observation is expected to consist of a pair of
images taken in the same bandpass within 30--60~minutes of each other,
in order to identify asteroids, and we count these as separate epochs.
The expected Medium Deep Survey cadence is to simply rotate through
the five filters every four days.

Tables~\ref{tbl:3pi} and \ref{tbl:m-d} summarize the relevant
information for each survey, as well as detection and saturation
limits that we derived. For the detection limits, we have assumed an
aperture that is 1 FWHM in diameter, which for a 2-dimensional
Gaussian collects 50\% of the source flux. We also tabulate the
earliest spectral type that is unsaturated at 100~pc, which is the
limiting distance at which PS1 is able to measure S/N~=~10
parallaxes. Unless the seeing is worse than the nominal 0\farcs8 (see
Section~\ref{sec:assump}), objects earlier than this will not have
measurable parallaxes because they will be saturated.  Loss of time
due to weather has been accounted for in the survey parameters.

Using the survey information given in Tables~\ref{tbl:3pi} and
\ref{tbl:m-d} and the empirical relations in
Section~\ref{sec:sim-star}, we have derived detection limits as a
function of spectral type for each PS1 bandpass.
Figure~\ref{fig:survey-limits} shows the distances at which objects of
a given spectral type are expected to reach the detection threshold
(S/N~=~5), as well as the distances within which they are so bright
that they will be saturated. Because the color selection of nearby
stars may benefit from, or require, the use of near-infrared
photometry, we also show the distance limit at which 2MASS is 99\%
complete in $J$ band ($J$~$<$~15.8~mag). We chose not to show the
$K$-band limit for simplicity, although it is very similar to the
$J$-band limit (except for T dwarfs) due to the trade off between a
brighter sky background but more object flux for red objects at $K$
band.

\clearpage
\section{Algorithms for Detecting Eclipses in Sparsely Sampled Light Curves\label{app:alg}}

We have investigated algorithms for detecting eclipse signatures in
sparsely sampled light curves, such as those generated by upcoming
multi-epoch all-sky surveys.  This problem is inherently different
from searching for a low-amplitude periodic signal in a densely
sampled light curve, for which robust methods have been developed for
current planetary transit surveys \citep[e.g,][]{2002A&A...391..369K}.
Rather, a sparsely sampled light curve will likely only catch one
eclipse event at a single epoch, so we are interested in identifying
a single discrepant point for an otherwise constant star.  Below we
describe three algorithms for identifying such eclipses and then
assess their detection limits and vulnerability to false positives.

\begin{enumerate}

\item \emph{$\chi^2$~null test}. The $\chi^2$ of a light curve may be
  reliably computed when measurement errors are well determined:
  \begin{eqnarray} \label{eqn:chi2}
    \chi^2 & = & \sum_{i=1}^n \frac{(N_i - \overline{N})^2}{\sigma_i^2}
  \end{eqnarray}
  where $N_i$ and $\sigma_i$ are the counts (DN) and uncertainty at
  each light curve point, and $\overline{N}$ is the weighted average
  of the light curve. Any deviation from a constant light curve will
  increase the value of $\chi^2$. For example, an increase in $\chi^2$
  could signify a single very discrepant point (e.g., a deep eclipse)
  or variability throughout the light curve.  The value of $\chi^2$
  gives the confidence interval for the null hypothesis that the flux
  from the star is constant.

\item \emph{$S^2$~index}. \citet{2005MNRAS.359.1096B} introduced an
  eclipse detection statistic ($S^2$) which combines the $\chi^2$
  values of different portions of the light curve to give an index
  which is roughly equivalent to the (S/N)$^2$ of the detection.  We
  adopt the form of the $S^2$ index as modified by
  \citet{2008MNRAS.386..887B}, which was used to assess the presence
  of eclipses and flares in sparsely sampled light curves in SDSS
  Stripe 82:
  \begin{eqnarray} \label{eqn:sind}
    S^2 & = & \frac{\chi^2_{\rm const} - \chi^2_{\rm out}}{\chi^2_{\rm out}/\nu_{\rm const}}
  \end{eqnarray}
  where $\chi^2_{\rm const}$ is the same as Equation~\ref{eqn:chi2},
  $\chi^2_{\rm out}$ is computed excluding the minimum point in the
  light curve, and $\nu_{\rm const}$ is the number of
  degrees of freedom of $\chi^2_{\rm const}$ (i.e., $n-1$ for a light
  curve with $n$ points). The $S^2$ index is designed to be more
  robust against false positives due to variability than the $\chi^2$
  null tests, since $S^2$ is only large when the out-of-transit
  portion of the light curve is nearly constant (small $\chi^2_{\rm
    out}$). To determine whether an eclipse was detected, a threshold
  value of $S^2$ must be chosen (e.g., a $S^2$ value of 100 corresponds
  to a S/N of $\sim$10).

\item \emph{S/N model}. The S/N of a candidate eclipse event can be
  computed directly from the light curve and its errors:
  \begin{eqnarray} \label{eqn:sntr}
    \frac{\rm S}{\rm N} & = & \frac{\overline{N}_{\rm out} - N_{\rm min}}{\sqrt{\sigma_{\rm out}^2 + \sigma_{\rm min}^2}}
  \end{eqnarray}
  where $N_{\rm min}$ and $\sigma_{\rm min}$ are the counts (DN) and
  error at light curve's minimum point, $\overline{N}_{\rm out}$ is
  the mean of the remaining points, and $\sigma_{\rm out}$ is the
  standard error of the mean. In our S/N model, $\sigma_{\rm out}$ and
  $\sigma_{\rm min}$ are limited by the systematic noise floor
  (S/N$_{\rm floor}$~=~100 for PS1).  For example, the uncertainty in
  the out-of-transit counts ($\sigma_{\rm out}$) is not allowed to be
  smaller than $\overline{N}_{\rm out}$/(S/N$_{\rm floor}$).  The
  eclipse S/N computed using this model will be small for a constant
  light curve and large when the minimum point is very different from
  the remaining points.  A threshold S/N must be chosen to determine
  whether an eclipse was detected.

\end{enumerate}

\begin{figure} 
\centerline{\includegraphics[width=3.5in,angle=0]{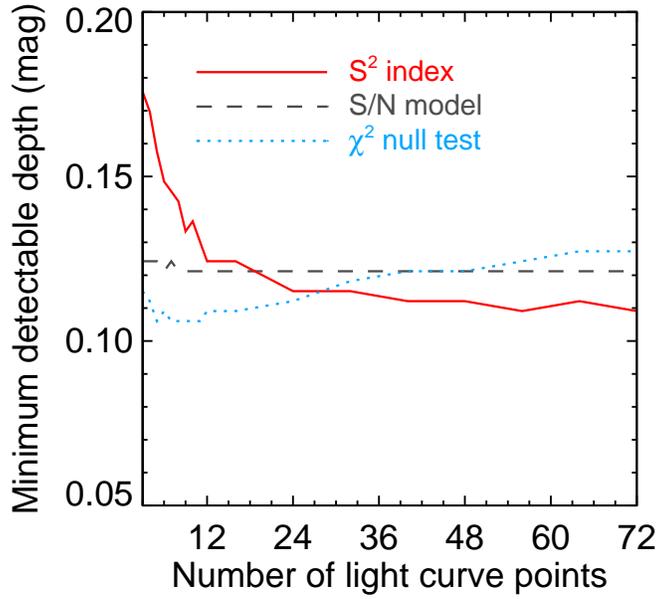}}

\caption{ \normalsize Minimum detectable eclipse depths for a source
  with S/N~=~100 per epoch using the algorithms described in the text,
  adopting an eclipse threshold S/N of 7.  The $S^2$ index
  \citep[solid line,][]{2008MNRAS.386..887B} improves significantly as
  more light curve points are added.  The S/N model (dashed line) does
  not improve because we have assumed a photometric floor at
  S/N~=~100, so adding more light curve points does not reduce the
  noise in this example. The $\chi^2$ null test (dotted line) becomes
  less sensitive with the number of light curve points, since adding
  more constant flux points reduces the significance of the single
  discrepant point. \label{fig:alg-mindepth}}

\end{figure}

The ideal eclipse detection algorithm is sensitive to the smallest
flux diminutions, while also being robust against false positives. In
Figure~\ref{fig:alg-mindepth}, we show the minimum detectable eclipse
depths of the various algorithms for a source with a photometric S/N
of 100 (i.e., at the photometric error floor of PS1). As points are
added to the light curve, the sensitivity of the algorithms change in
different ways. For the $S^2$ index, the sensitivity monotonically
improves, while for the $\chi^2$ null test the sensitivity actually
gets worse. This is simply due to the fact that by adding more
constant flux points to the light curve, the significance of the
single discrepant point is decreased compared to the null hypothesis
of constant flux. The S/N model gives essentially the same sensitivity
for any light curve because in this case we have considered a source
whose photometric errors are already at the noise floor; thus, adding
more points does not reduce the noise. In
Figure~\ref{fig:alg-mindepth}, we have chosen a $S^2$ threshold of 150
in order to roughly match the sensitivity of the S/N model threshold
of 10 and $\chi^2$ null test confidence limit of 10$\sigma$. Of
course, the exact choice of the detection threshold should be based on
the expected number of false positives.

\begin{figure} 
\centerline{\includegraphics[width=6.5in,angle=0]{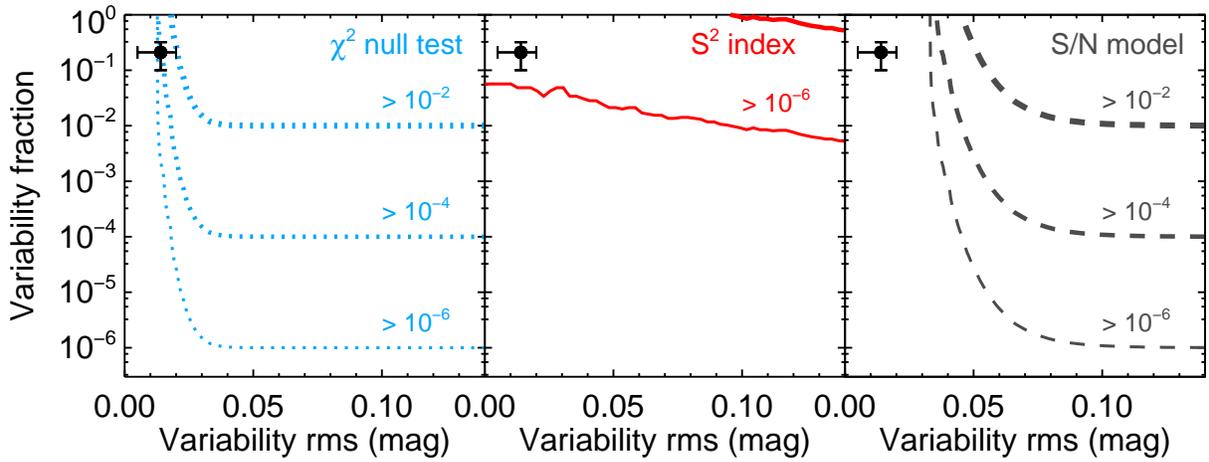}}

\caption{ \normalsize Contours of the false positive rate due to star
  spots for the three eclipse detection algorithms described in the
  text.  Contours are drawn at false positive rates of 10$^{-2}$,
  10$^{-4}$, and 10$^{-6}$, adopting an eclipse threshold S/N of 7. We
  modeled star spots as sinusoidal variability with over a range
  amplitudes (0.0~$<$~rms~$<$~0.14~mag).  As a point of comparison,
  the best available estimate of variability among M~dwarfs from
  \citet{2006MNRAS.367..407R} is shown as a filled circle with errors.
  The $\chi^2$ null test (dotted line) is the most vulnerable to
  contamination from star spots, while the $S^2$ index (solid line)
  and S/N model (dashed line) are vulnerable to more limited,
  different source populations of false positives: rare, highly
  variable objects for the S/N model and common objects with a low
  level of variability for the $S^2$ index. \label{fig:alg-false-pos}}

\end{figure}

In practice, the number of false positives resulting from different
detection algorithms depends on the particular survey strategy and the
nature of the contaminating sources. In our simulations, we have found
that PS1 will be particularly sensitive to transiting Jupiters around
M~dwarfs. A large fraction of M~dwarfs are chromospherically active
\citep[e.g.,][]{2000AJ....120.1085G}, thus they are expected to show
photometric variability due to spots, making this one of the most
likely sources of false positive eclipse detections in sparsely
sampled data. In Figure~\ref{fig:alg-false-pos}, we show contours of
the false positive fraction due to spots for the 3$\pi$~Survey's
12-point light curves. The variability due to spots is assumed to be
sinusoidal, and we have considered a range of amplitudes, properly
accounting for the fact that a sine curve has an rms of
$\sin(\frac{\pi}{4})$. We adopt the best available estimate of
variability among M~dwarfs from \citet{2006MNRAS.367..407R} who found
a median variability amplitude of 0.014~mag (0.005--0.02~mag observed
range) for 21$\pm$11\% of M~dwarfs. The contours in
Figure~\ref{fig:alg-false-pos} show the fraction of false positives
over a range of parameter space that covers low to high amplitude
variability and rare to common occurence fractions of variability.

For the 3$\pi$~Survey, Figure~\ref{fig:alg-false-pos} shows that the
$\chi^2$ null test is vulnerable to false positives over a broader
range of variability parameter space than the other algorithms.  This
is intuitively expected as it is sensitive to any deviations from a
constant light curve, not just eclipses.  The $S^2$ index and S/N
model yield false positives over about the same amount of parameter
space, but in different regions.  The S/N model is vulnerable to very
rare, highly variable objects.  For example, if 1 in 10$^3$ M~dwarfs
are variable at the 0.10~mag level, this would result in an
unmanageable false positive fraction of $\sim$10$^{-3}$.  Such a
population could easily have been missed by the best available
variability surveys, which have targeted only a few dozen M~dwarfs.
On the other hand, the $S^2$ index is vulnerable to relatively common
objects with a lower level of variability.  We also experimented with
arbitrarily increasing the number of light curve points and found the
most drastic improvement for the $S^2$ index which for light curves
with $>$~15~points had a false positive rate $<$~10$^{-6}$ over the
entire variability parameter space in
Figure~\ref{fig:alg-false-pos}. The improvement for the S/N model was
less extreme, with contours changing by only 0.02~mag for 72-point
light curves, and the $\chi^2$ null test actually worsened.

In summary, we have investigated three algorithms for eclipse
detection in sparsely sampled light curves with very few ($\sim$10)
epochs. We have paid particular attention to possible contamination
from variability due to star spots, which is the most likely source of
eclipse false positives in such data for M~dwarfs. All algorithms
achieve similar eclipse depth sensitivities for a given confidence
interval (Figure~\ref{fig:alg-mindepth}). We found that the $\chi^2$
null test is a good general tool for identifying deviations from a
constant light curve, but because of this it has the highest false
positive rate. The $S^2$ index and S/N model are vulnerable to
different source populations of false positives (e.g., rare, highly
variable objects), which are largely unprobed by current variability
surveys. Thus, it is impossible to predict the better algorithm for
robustly detecting transits in advance of actual survey data. In fact,
using \emph{both} algorithms simultaneously may prove to be the best
strategy for eliminating false positives. We emphasize that the proper
eclipse detection algorithm depends not only on the underlying source
of false positives, but also on the particular survey strategy. For
example, in the PS1 3$\pi$~Survey observation epochs will actually be
obtained in pairs separated by 30--60~minutes for the purpose of
asteroid detection. This may help distinguish spots from true eclipse
events since stellar rotation for M~dwarfs occurs over much longer
time scales \citep[$P$~$\gtrsim$~1~day, e.g.,][]{1998A&A...331..581D},
and thus spots would cause a flux dip in both images in the pair,
whereas a transit would only be detected in one \citep[e.g., see the
discussion on paired observations by][]{1993AJ....105.1813W}.



\begin{figure} 
\centerline{\includegraphics[width=6.5in,angle=0]{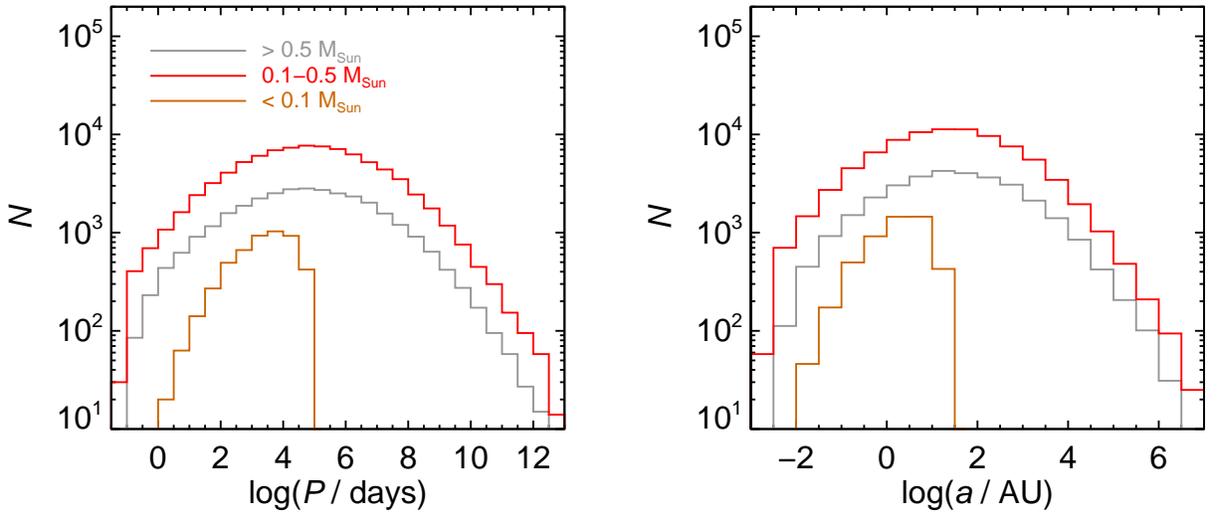}}

\caption{ \normalsize Orbital period and semimajor axis distributions
  of simulated binary companions in the PS1 3$\pi$~Survey.  For FGKM
  stars ($>$~0.1~\Msun), we drew periods from the DM91 log-normal
  period distribution.  For very low-mass stars and brown dwarfs
  ($<$~0.1~\Msun), we drew periods from the log-normal period
  distribution of \citet{2006AJ....132..663B}, which is truncated at
  15~AU.  We converted period into semimajor axis using the system
  mass $M_1(1+q)$, where $q$ is the mass ratio.  The distributions
  appear truncated at short periods because 0.6\% of companions were
  rejected on the basis of being non-detached. \label{fig:per-sep}}

\end{figure}

\begin{figure} 
\centerline{\includegraphics[width=6.5in,angle=0]{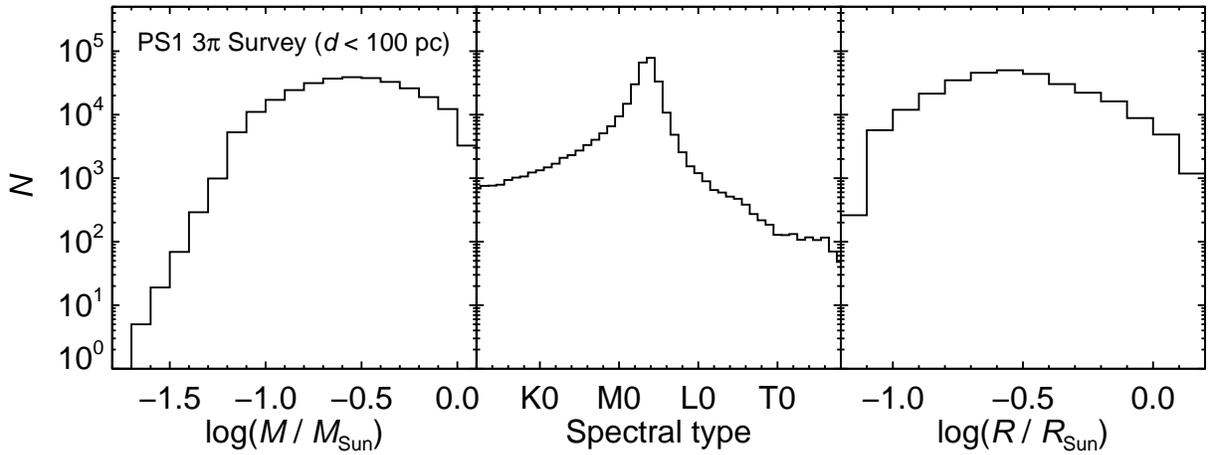}}

\caption{ \normalsize Predicted properties of objects detected in the
  PS1 3$\pi$~Survey, restricted to stars within 100~pc for which PS1
  will measure parallaxes. We have adopted the log-normal mass
  function of \citet{2008AJ....136.1778C}, which appears truncated on
  the high mass end here due to the saturation limit of the 3$\pi$
  Survey.  The spectral type distribution shows that, by far, the most
  common objects in this volume-limited sample are mid-M~dwarfs. PS1
  will also detect hundreds of T dwarfs and thousands of L dwarfs in
  the solar neighborhood. As a consequence of the preeminence of mid-M
  dwarfs, the typical radius in this sample is $\sim$0.3~\Rsun.  For
  such objects, transiting Jupiters would cause a $\sim$10\% flux
  diminution. \label{fig:sim-star}}

\end{figure}

\begin{figure} 
\centerline{\includegraphics[width=3.5in,angle=0]{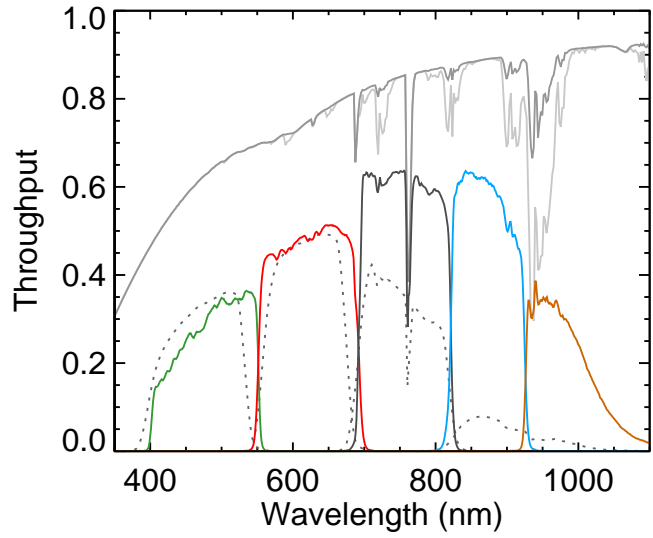}}

\caption{ \normalsize The total throughput of Pan-STARRS-1 bandpasses
  $g$, $r$, $i$, $z$, and $y$ (green, red, dark gray, blue, and brown,
  respectively).  For comparison, the SDSS bands are shown (dotted
  gray lines), with the primary differences being the absence of an
  SDSS $y$ band, the longpass SDSS $z$ band, and the greater overall
  red sensitivity of PS1 due to improved detector quantum efficiency
  at redder wavelengths.  The atmospheric throughput at an airmass of
  2 is also shown for 0.8~mm~H$_2$O (darker gray) and 8~mm~H$_2$O
  (lighter gray). \label{fig:ps1-filt}}

\end{figure}

\clearpage
\begin{deluxetable}{lcccccc}
\tablewidth{0pt}
\tablecaption{ Model of Extended Solar Neighborhood ($d$~$<$~100~pc) \label{tbl:100pc}}
\tablehead{
\colhead{Spectral Type}   &
\colhead{$N_{\rm all}$\tablenotemark{a}}   &
\colhead{PS1 3$\pi$ (saturated)\tablenotemark{b}}}

\startdata

                         \multicolumn{3}{c}{Stars and brown dwarfs} \\
                         \cline{1-3}
Total                 &   4.0$\times$10$^{  5}$  &   2.1$\times$10$^{  5}$ (8.5$\times$10$^{  4}$) \\ 
G+K dwarfs            &   5.2$\times$10$^{  4}$  &   0 \phn\phn\phn\phn\phn\phn(3.9$\times$10$^{  4}$) \\ 
M~dwarfs              &   3.4$\times$10$^{  5}$  &   2.1$\times$10$^{  5}$ (4.6$\times$10$^{  4}$) \\ 
L dwarfs              &                    8000  &                    5400 (0)\phn\phn\phn \\                   
T dwarfs              &                    6900  &                 \phn880 (0)\phn\phn\phn \\                   
                         \cline{1-3}

                         \multicolumn{3}{c}{} \\
                         \multicolumn{3}{c}{Transiting Jupiter hosts\tablenotemark{c}} \\
                         \cline{1-3}
Total                 &                 \phn142  &              \phn\phn57 (47)\phn\phn\phn \\                    
G+K dwarfs            &              \phn\phn44  &           \phn\phn\phn0 (33)\phn\phn\phn \\                    
M~dwarfs              &              \phn\phn94  &              \phn\phn56 (14)\phn\phn\phn \\                    
L+T dwarfs            &           \phn\phn\phn4  &           \phn\phn\phn1 (0)\phn\phn\phn\phn \\                                           \cline{1-3}
   
                         \multicolumn{3}{c}{} \\
                         \multicolumn{3}{c}{Eclipsing binary primaries\tablenotemark{c}} \\
                         \cline{1-3}
Total                 &                    1284  &                 \phn573 (400)\phn\phn\phn \\                   
G+K dwarfs            &                 \phn280  &           \phn\phn\phn0 (210)\phn\phn\phn \\                   
M~dwarfs              &                    1000  &                 \phn570 (190)\phn\phn\phn \\                   
L+T dwarfs            &           \phn\phn\phn4  &           \phn\phn\phn3 (0)\phn\phn\phn\phn\phn \\                     

\enddata

\tablenotetext{a}{All objects within 100~pc.}

\tablenotetext{b}{Number of sources within the magnitude limits of
  (saturated in) the Pan-STARRS 3$\pi$~Survey.}

\tablenotetext{c}{Number of systems with orbits aligned such that
  transits/eclipses are allowed, but not necessarily detected by the
  PS1 3$\pi$~Survey. These numbers reflect how many such systems are
  within the magnitude limits of the 3$\pi$~Survey.}

\end{deluxetable}

\clearpage
\begin{deluxetable}{cll}
\tablewidth{0pt}
\tablecaption{
Empirical Spectral Type Relations \label{tbl:spt-rel}}
\tablehead{
\colhead{Poly. fit} &
\colhead{SpT -- \Lbol} &
\colhead{BC$_K$ -- SpT}}
\startdata
$c_0$  &   $-$1.5917$\times$10$^{  1}$  &  \phs2.3785                   \\
$c_1$  &   $-$1.8829$\times$10$^{  1}$  &  \phs9.3034$\times$10$^{ -2}$ \\
$c_2$  &   $-$3.9862                    &  \phs1.2667$\times$10$^{ -3}$ \\
$c_3$  &  \phs1.3840                    &   $-$1.9272$\times$10$^{ -4}$ \\
$c_4$  &  \phs6.6571$\times$10$^{ -1}$  &   $-$4.9479$\times$10$^{ -6}$ \\
$c_5$  &  \phs6.3156$\times$10$^{ -2}$  &  \phs1.9198$\times$10$^{ -7}$ \\
rms    &  \phs1.1                       & \phs0.08                     \\
\enddata

\tablerefs{Polynomial fits are valid between spectral types G0 and T8,
  where
$${\rm SpT} = \sum_{i=0}^5 c_i \times \log(\Lbol/\Lsun)^{i}$$
$${\rm BC}_K = \sum_{i=0}^5 c_i \times {\rm SpT}^{i}$$
Numerical spectral types are defined as (G0, K0, M0, L0, T0)~=~($-$20,
$-$10, 0, 10, 20).}

\end{deluxetable}

\clearpage
\begin{deluxetable}{ccccccc}
\tablewidth{0pt}
\tablecaption{
Optical and Near-infrared Colors of Low-Mass Stars and Brown Dwarfs\tablenotemark{a} \label{tbl:spt-color}}
\tablehead{
\colhead{SpT} &
\colhead{$g-r$} &
\colhead{$r-i$} &
\colhead{$i-z$} &
\colhead{$z-J$} &
\colhead{$y-J$} &
\colhead{$J-K$}}
\startdata
G0 &  0.32$\pm$0.10                  &  0.14$\pm$0.10                  &  0.02$\pm$0.10                  &  0.73$\pm$0.10                  &  0.75$\pm$0.10                  & \phs0.24$\pm$0.10                  \\
G1 &  0.35$\pm$0.10\tablenotemark{b} &  0.14$\pm$0.10\tablenotemark{b} &  0.03$\pm$0.10\tablenotemark{b} &  0.81$\pm$0.10\tablenotemark{b} &  0.80$\pm$0.10\tablenotemark{b} & \phs0.27$\pm$0.10\tablenotemark{b} \\
G2 &  0.38$\pm$0.10                  &  0.15$\pm$0.10                  &  0.04$\pm$0.10                  &  0.88$\pm$0.10                  &  0.85$\pm$0.10                  & \phs0.30$\pm$0.10                  \\
G3 &  0.39$\pm$0.10\tablenotemark{b} &  0.15$\pm$0.10\tablenotemark{b} &  0.04$\pm$0.10\tablenotemark{b} &  0.86$\pm$0.10\tablenotemark{b} &  0.83$\pm$0.10\tablenotemark{b} & \phs0.29$\pm$0.10\tablenotemark{b} \\
G4 &  0.40$\pm$0.10\tablenotemark{b} &  0.16$\pm$0.10\tablenotemark{b} &  0.04$\pm$0.10\tablenotemark{b} &  0.84$\pm$0.10\tablenotemark{b} &  0.80$\pm$0.10\tablenotemark{b} & \phs0.28$\pm$0.10\tablenotemark{b} \\
G5 &  0.41$\pm$0.10                  &  0.16$\pm$0.10                  &  0.05$\pm$0.10                  &  0.82$\pm$0.10                  &  0.78$\pm$0.10                  & \phs0.27$\pm$0.10                  \\
G6 &  0.43$\pm$0.10\tablenotemark{b} &  0.17$\pm$0.10\tablenotemark{b} &  0.05$\pm$0.10\tablenotemark{b} &  0.85$\pm$0.10\tablenotemark{b} &  0.82$\pm$0.10\tablenotemark{b} & \phs0.30$\pm$0.10\tablenotemark{b} \\
G7 &  0.46$\pm$0.10\tablenotemark{b} &  0.17$\pm$0.10\tablenotemark{b} &  0.05$\pm$0.10\tablenotemark{b} &  0.87$\pm$0.10\tablenotemark{b} &  0.86$\pm$0.10\tablenotemark{b} & \phs0.33$\pm$0.10\tablenotemark{b} \\
G8 &  0.49$\pm$0.10                  &  0.18$\pm$0.10                  &  0.06$\pm$0.10                  &  0.90$\pm$0.10                  &  0.90$\pm$0.10                  & \phs0.36$\pm$0.10                  \\
G9 &  0.51$\pm$0.10\tablenotemark{b} &  0.20$\pm$0.10\tablenotemark{b} &  0.06$\pm$0.10\tablenotemark{b} &  0.89$\pm$0.10\tablenotemark{b} &  0.88$\pm$0.10\tablenotemark{b} & \phs0.41$\pm$0.10\tablenotemark{b} \\
K0 &  0.53$\pm$0.10                  &  0.21$\pm$0.10                  &  0.06$\pm$0.10                  &  0.88$\pm$0.10                  &  0.85$\pm$0.10                  & \phs0.47$\pm$0.10                  \\
K1 &  0.60$\pm$0.10\tablenotemark{b} &  0.22$\pm$0.10\tablenotemark{b} &  0.08$\pm$0.10\tablenotemark{b} &  0.96$\pm$0.10\tablenotemark{b} &  0.92$\pm$0.10\tablenotemark{b} & \phs0.49$\pm$0.10\tablenotemark{b} \\
K2 &  0.66$\pm$0.10                  &  0.23$\pm$0.10                  &  0.10$\pm$0.10                  &  1.03$\pm$0.10                  &  0.99$\pm$0.10                  & \phs0.52$\pm$0.10                  \\
K3 &  0.74$\pm$0.10                  &  0.31$\pm$0.10                  &  0.11$\pm$0.10                  &  1.04$\pm$0.10                  &  1.00$\pm$0.10                  & \phs0.52$\pm$0.10                  \\
K4 &  0.86$\pm$0.10                  &  0.37$\pm$0.10                  &  0.15$\pm$0.10                  &  1.06$\pm$0.10                  &  1.00$\pm$0.10                  & \phs0.62$\pm$0.10                  \\
K5 &  1.02$\pm$0.10                  &  0.41$\pm$0.10                  &  0.16$\pm$0.10                  &  1.12$\pm$0.10                  &  1.02$\pm$0.10                  & \phs0.64$\pm$0.10                  \\
K6 &  1.08$\pm$0.10\tablenotemark{b} &  0.48$\pm$0.10\tablenotemark{b} &  0.21$\pm$0.10\tablenotemark{b} &  1.10$\pm$0.10\tablenotemark{b} &  0.98$\pm$0.10\tablenotemark{b} & \phs0.65$\pm$0.10\tablenotemark{b} \\
K7 &  1.14$\pm$0.10                  &  0.55$\pm$0.10                  &  0.25$\pm$0.10                  &  1.07$\pm$0.10                  &  0.95$\pm$0.10                  & \phs0.65$\pm$0.10                  \\
K8 &  1.16$\pm$0.25\tablenotemark{b} &  0.59$\pm$0.11\tablenotemark{b} &  0.27$\pm$0.09\tablenotemark{b} &  1.14$\pm$0.13\tablenotemark{b} &  1.00$\pm$0.10\tablenotemark{b} & \phs0.71$\pm$0.11\tablenotemark{b} \\
K9 &  1.19$\pm$0.39\tablenotemark{b} &  0.63$\pm$0.11\tablenotemark{b} &  0.28$\pm$0.08\tablenotemark{b} &  1.22$\pm$0.16\tablenotemark{b} &  1.05$\pm$0.10\tablenotemark{b} & \phs0.77$\pm$0.13\tablenotemark{b} \\
M0 &  1.21$\pm$0.54                  &  0.67$\pm$0.12                  &  0.30$\pm$0.07                  &  1.29$\pm$0.19                  &  1.10$\pm$0.10\tablenotemark{b} & \phs0.83$\pm$0.14                  \\
M1 &  1.27$\pm$0.44                  &  0.83$\pm$0.18                  &  0.36$\pm$0.07                  &  1.35$\pm$0.21                  &  1.15$\pm$0.09                  & \phs0.85$\pm$0.13                  \\
M2 &  1.39$\pm$0.35                  &  1.01$\pm$0.13                  &  0.44$\pm$0.08                  &  1.40$\pm$0.17                  &  1.15$\pm$0.07\tablenotemark{c} & \phs0.84$\pm$0.12                  \\
M3 &  1.38$\pm$0.22                  &  1.21$\pm$0.16                  &  0.52$\pm$0.08                  &  1.48$\pm$0.17                  &  1.16$\pm$0.04                  & \phs0.84$\pm$0.14                  \\
M4 &  1.34$\pm$0.20                  &  1.46$\pm$0.15                  &  0.64$\pm$0.09                  &  1.57$\pm$0.16                  &  1.26$\pm$0.08                  & \phs0.86$\pm$0.13                  \\
M5 &  1.32$\pm$0.27                  &  1.90$\pm$0.13                  &  0.89$\pm$0.07                  &  1.75$\pm$0.11                  &  1.41$\pm$0.19                  & \phs0.90$\pm$0.14                  \\
M6 &  1.34$\pm$0.13                  &  2.08$\pm$0.14                  &  0.97$\pm$0.07                  &  1.87$\pm$0.10                  &  1.56$\pm$0.18                  & \phs0.93$\pm$0.14                  \\
M7 &  1.39$\pm$0.17                  &  2.46$\pm$0.18                  &  1.16$\pm$0.09                  &  2.03$\pm$0.10                  &  1.62$\pm$0.17\tablenotemark{c} & \phs0.98$\pm$0.11                  \\
M8 &  1.60$\pm$0.16                  &  2.67$\pm$0.25                  &  1.35$\pm$0.15                  &  2.24$\pm$0.12                  &  1.70$\pm$0.16\tablenotemark{b} & \phs1.06$\pm$0.06                  \\
M9 &  1.56$\pm$0.14                  &  2.73$\pm$0.21                  &  1.51$\pm$0.08                  &  2.48$\pm$0.09                  &  1.77$\pm$0.15\tablenotemark{c} & \phs1.12$\pm$0.08                  \\
L0 &  2.34$\pm$0.39                  &  2.38$\pm$0.09                  &  1.57$\pm$0.06                  &  2.70$\pm$0.11                  &  1.90$\pm$0.13\tablenotemark{b} & \phs1.27$\pm$0.06                  \\
L1 &  \nodata                        &  2.43$\pm$0.27                  &  1.62$\pm$0.09                  &  2.81$\pm$0.08                  &  2.03$\pm$0.12\tablenotemark{b} & \phs1.28$\pm$0.19                  \\
L2 &  \nodata                        &  2.36$\pm$0.26                  &  1.58$\pm$0.10                  &  2.80$\pm$0.10                  &  2.16$\pm$0.11\tablenotemark{b} & \phs1.43$\pm$0.20                  \\
L3 &  \nodata                        &  2.06$\pm$0.18                  &  1.59$\pm$0.17                  &  3.15$\pm$0.16                  &  2.29$\pm$0.10                  & \phs1.56$\pm$0.19                  \\
L4 &  \nodata                        &  \nodata                        &  1.79$\pm$0.29                  &  3.30$\pm$0.16                  &  2.18$\pm$0.12                  & \phs1.64$\pm$0.20                  \\
L5 &  \nodata                        &  \nodata                        &  1.92$\pm$0.16                  &  3.22$\pm$0.15                  &  2.15$\pm$0.09                  & \phs1.65$\pm$0.20                  \\
L6 &  \nodata                        &  \nodata                        &  2.06$\pm$0.33                  &  3.34$\pm$0.13                  &  2.11$\pm$0.10                  & \phs1.65$\pm$0.20                  \\
L7 &  \nodata                        &  \nodata                        &  2.31$\pm$0.24                  &  3.33$\pm$0.13                  &  2.05$\pm$0.11\tablenotemark{c} & \phs1.69$\pm$0.20                  \\
L8 &  \nodata                        &  \nodata                        &  2.26$\pm$0.31                  &  3.38$\pm$0.10                  &  2.10$\pm$0.12                  & \phs1.63$\pm$0.17                  \\
L9 &  \nodata                        &  \nodata                        &  2.31$\pm$0.40\tablenotemark{b} &  3.34$\pm$0.07                  &  2.06$\pm$0.02                  & \phs1.53$\pm$0.09                  \\
T0 &  \nodata                        &  \nodata                        &  2.36$\pm$0.49                  &  3.47$\pm$0.21                  &  2.14$\pm$0.10\tablenotemark{c} & \phs1.44$\pm$0.20                  \\
T1 &  \nodata                        &  \nodata                        &  \nodata                        &  3.75$\pm$0.21                  &  2.18$\pm$0.17                  & \phs1.09$\pm$0.20                  \\
T2 &  \nodata                        &  \nodata                        &  \nodata                        &  3.83$\pm$0.23                  &  2.22$\pm$0.11                  & \phs0.77$\pm$0.20                  \\
T3 &  \nodata                        &  \nodata                        &  \nodata                        &  4.05$\pm$0.06                  &  2.23$\pm$0.10\tablenotemark{c} & \phs0.34$\pm$0.20                  \\
T4 &  \nodata                        &  \nodata                        &  \nodata                        &  4.28$\pm$0.06\tablenotemark{b} &  2.39$\pm$0.09\tablenotemark{c} & $-$0.07$\pm$0.20                  \\
T5 &  \nodata                        &  \nodata                        &  \nodata                        &  4.50$\pm$0.06\tablenotemark{b} &  2.41$\pm$0.08                  & $-$0.30$\pm$0.19                  \\
T6 &  \nodata                        &  \nodata                        &  \nodata                        &  4.73$\pm$0.07                  &  2.45$\pm$0.06\tablenotemark{b} & $-$0.71$\pm$0.20                  \\
T7 &  \nodata                        &  \nodata                        &  \nodata                        &  \nodata                        &  2.49$\pm$0.05\tablenotemark{b} & $-$0.53$\pm$0.20                  \\
T8 &  \nodata                        &  \nodata                        &  \nodata                        &  \nodata                        &  2.53$\pm$0.04\tablenotemark{c} & $-$0.58$\pm$0.20                  \\
\enddata

\tablenotetext{a}{Optical photometry ($grizy$) on the PS1 system in AB
  magnitudes.  Near-infrared photometry ($JK$) on the MKO system in
  Vega magnitudes.}

\tablenotetext{b}{Mean and rms are interpolated from neighboring
  spectral type bins.}

\tablenotetext{c}{Only the rms is interpolated from neighboring
  spectral type bins.}

\tablerefs{ \citet{1998PASP..110..863P, 2002astro.ph..4065H,
    2004AJ....127.3553K, 2004PASP..116....9S, chiu05,
    2007AJ....133..531B, 2008AJ....135..785W, 2009AJ....137....1F} }

\end{deluxetable}

\clearpage
\begin{deluxetable}{lcccccc}
\tablewidth{0pt}
\tablecaption{ PS1 3$\pi$ Simulation Results for Number of Eclipsing Systems\label{tbl:sim-res}}
\tablehead{
\colhead{}   &
\multicolumn{5}{c}{Bandpass}      &
\colhead{} \\
\cline{2-6}
\colhead{Survey Yield}    &
\colhead{$g$}   &
\colhead{$r$}   &
\colhead{$i$}   &
\colhead{$z$}   &
\colhead{$y$}   &
\colhead{~~~Total (F.P.)\tablenotemark{a}}}

\startdata
                         & \multicolumn{6}{c}{Semi-analytic} \\
                           \cline{2-7}

Transiting Jupiters & 3.7 & 3.5 & 2.8 & 2.6 & 3.3 & 15.9\tablenotemark{b}\phs\phn\phn\phn \\

                           \cline{2-7}
                           \multicolumn{6}{c}{} \\
                         & \multicolumn{6}{c}{Monte Carlo} \\
                           \cline{2-7}

Transiting Jupiters & 3.4 & 3.5 & 2.9 & 2.8 & 3.4 & 12.6\tablenotemark{c} (0.0)   \\ 
Eclipsing binaries  & 160 & 140 &  80 &  90 & 110 &  260\tablenotemark{c} (0)\phn \\ 

                           \cline{2-7}
                           \multicolumn{6}{c}{} \\
\enddata

\tablenotetext{a}{Number of false positives is given in parentheses.}

\tablenotetext{b}{This total does not account for double counting
  caused by transits detected in multiple bandpasses around the same
  star.}

\tablenotetext{c}{This total counts transits detected in multiple
  bandpasses for the same star only once, and thus is not simply the
  sum of the number of detections in individual bandpasses.}

\end{deluxetable}

\clearpage
\begin{deluxetable}{ccccccc}
\tablewidth{0pt}
\tablecaption{
AB Zero-Magnitude Properties of PS1 Bandpasses \label{tbl:ps1-filt}}
\tablehead{
\colhead{Band} &
\colhead{$\lambda_{\rm{eff},0}$\tablenotemark{a}} &
\colhead{$\Phi_0$\tablenotemark{b}} &
\colhead{$F_{\lambda,0}$\tablenotemark{a}} &
\colhead{$\Phi_{\rm sky}$\tablenotemark{c}} &
\colhead{$\Delta_{\rm Vega}$\tablenotemark{d}} \\
\colhead{} &
\colhead{\micron} &
\colhead{DN s$^{-1}$ m$^{-2}$} &
\colhead{W m$^{-2}$ \micron$^{-1}$} &
\colhead{DN s$^{-1}$ m$^{-2}$ \arcsec$^{-2}$} &
\colhead{mag}}
\startdata
$g$  &  0.484  &  4.73$\times$10$^{  9}$  &  4.75$\times$10$^{ -8}$  &\phn6.9  & \phs0.08  \\
$r$  &  0.621  &  5.87$\times$10$^{  9}$  &  2.86$\times$10$^{ -8}$  &   22.5  &  $-$0.16  \\
$i$  &  0.754  &  5.55$\times$10$^{  9}$  &  1.93$\times$10$^{ -8}$  &   48.7  &  $-$0.38  \\
$z$  &  0.869  &  3.78$\times$10$^{  9}$  &  1.45$\times$10$^{ -8}$  &   77.6  &  $-$0.54  \\
$y$  &  0.979  &  1.85$\times$10$^{  9}$  &  1.14$\times$10$^{ -8}$  &   89.6  &  $-$0.62  \\
\enddata

\tablenotetext{a}{Effective wavelength ($\lambda_{\rm{eff},0}$) and
  effective flux density ($F_{\lambda,0}$) in each bandpass for a
  zero-magnitude source. Computed by integrating the wavelength and
  flux density weighted by the transmission curve of the bandpass,
  $T(\lambda)$, and the photon flux density of the zero-magnitude
  source, $F_{\lambda,0}(\lambda)/(\frac{hc}{\lambda})$.}

\tablenotetext{b}{Total integrated photon flux of the zero-magnitude
  source in each bandpass.}

\tablenotetext{c}{Sky brightness derived from synthetic photometry of
  the Mauna Kea airglow spectrum obtained by
  \citet{1999Ap&SS.269..209S} using Keck/LRIS.  This sky spectrum
  terminates at 1.04~\micron, and since 8\% of the area of the $y$
  band lies beyond 1.04~\micron, the sky flux in this band is likely
  underestimated.}

\tablenotetext{d}{AB to Vega conversions ($\Delta_{\rm
    Vega}$~$\equiv$~$m_{\rm Vega}-m_{\rm AB}$) derived from synthetic
  photometry of a flux-calibrated spectrum of Vega
  \citep{1985IAUS..111..225H, 1985A&A...151..399M}.}

\end{deluxetable}

\clearpage
\begin{deluxetable}{cllll}
\tablewidth{0pt}
\tablecaption{
SDSS to PS1 Photometric System Conversions \label{tbl:phot-conv}}
\tablehead{
\colhead{Poly. fit} &
\colhead{$g_{\rm PS1}-g_{\rm SDSS}$} &
\colhead{$r_{\rm PS1}-r_{\rm SDSS}$} &
\colhead{$i_{\rm PS1}-i_{\rm SDSS}$} &
\colhead{$z_{\rm PS1}-z_{\rm SDSS}$}}
\startdata
$c_0$  &   $-$2.2409$\times$10$^{ -1}$  &   $-$1.5839$\times$10$^{ -2}$  &   $-$2.7831$\times$10$^{ -2}$  &  6.0387$\times$10$^{ -2}$  \\
$c_1$  &   $-$1.2085$\times$10$^{ -2}$  &   $-$4.3486$\times$10$^{ -3}$  &   $-$1.9926$\times$10$^{ -3}$  &  9.9246$\times$10$^{ -3}$  \\
$c_2$  &  \phs5.6155$\times$10$^{ -5}$  &   $-$5.9648$\times$10$^{ -4}$  &   $-$2.1100$\times$10$^{ -5}$  &  4.5737$\times$10$^{ -4}$  \\
$c_3$  &  \phs1.4624$\times$10$^{ -5}$  &   $-$2.1129$\times$10$^{ -5}$  &  \phs \nodata                 &  4.9887$\times$10$^{ -6}$  \\
\enddata

\tablerefs{Polynomial fits are a function of spectral type (SpT), where
$${\rm mag(PS1)}-{\rm mag(SDSS)} = \sum_{i=0}^5 c_i \times ({\rm SpT})^{i}$$
and numerical spectral types are defined as (G0, K0, M0, L0,
T0)~=~($-$20, $-$10, 0, 10, 20).  Fits are valid from G0 to T8 for $i$
and $z$ bands, from G0 to L5 for $r$ band, and from G0 to M6 for $g$
band.}

\end{deluxetable}

\clearpage
\begin{deluxetable}{ccccccc}
\tablewidth{0pt}
\tablecaption{
Pan-STARRS-1 3$\pi$~Survey \label{tbl:3pi}}
\tablehead{
\colhead{Band} &
\colhead{$N_{\rm epoch}$} &
\colhead{$t_{\rm exp}$} &
\colhead{$m_{\rm lim}$\tablenotemark{a}} &
\colhead{$m_{\rm floor}$\tablenotemark{b}} &
\colhead{$m_{\rm sat}$\tablenotemark{c}} &
\colhead{SpT$_{\rm sat}$\tablenotemark{d}} \\
\colhead{} &
\colhead{} &
\colhead{(s)} &
\colhead{(mag)} &
\colhead{(mag)} &
\colhead{(mag)} &
\colhead{}}
\startdata
 $g$  &  12  &   60  &  23.4  &  18.7  &  15.6  &  M1  \\
 $r$  &  12  &   38  &  22.8  &  18.3  &  15.3  &  M2  \\
 $i$  &  12  &   30  &  22.2  &  17.9  &  14.8  &  M3  \\
 $z$  &  12  &   30  &  21.6  &  17.4  &  14.4  &  M3  \\
 $y$  &  12  &   30  &  20.1  &  15.9  &  12.8  &  M2  \\
\enddata

\tablenotetext{a}{Faintest magnitude that has a S/N of 5.}

\tablenotetext{b}{Faintest magnitude that has a S/N of S/N$_{\rm floor}$~=~100.}

\tablenotetext{c}{Saturation limits derived for a Gaussian PSF with a
  FWHM of 0\farcs8, a gain of 1 $e^-$ DN$^{-1}$, a full-well depth of
  3.0$\times$10$^4$~$e^-$.}

\tablenotetext{d}{Earliest spectral type that is not saturated at
  100~pc.}

\end{deluxetable}

\clearpage
\begin{deluxetable}{ccccccc}
\tablewidth{0pt}
\tablecaption{
Pan-STARRS-1 Medium Deep Survey \label{tbl:m-d}}
\tablehead{
\colhead{Band} &
\colhead{$N_{\rm epoch}$} &
\colhead{$N \times t_{\rm exp}$} &
\colhead{$m_{\rm lim}$\tablenotemark{a}} &
\colhead{$m_{\rm floor}$\tablenotemark{b}} &
\colhead{$m_{\rm sat}$\tablenotemark{c}} &
\colhead{SpT$_{\rm sat}$\tablenotemark{d}} \\
\colhead{} &
\colhead{} &
\colhead{(s)} &
\colhead{(mag)} &
\colhead{(mag)} &
\colhead{(mag)} &
\colhead{}}
\startdata
 $g$  &  72  &  3$\times$240  &  24.8  &  20.9  &  16.8  &  M3  \\
 $r$  &  72  &  3$\times$240  &  24.4  &  20.8  &  16.8  &  M4  \\
 $i$  &  72  &  6$\times$240  &  24.4  &  20.9  &  16.4  &  M5  \\
 $z$  &  72  &  6$\times$240  &  23.7  &  20.3  &  15.7  &  M4  \\
 $y$  &  72  &  6$\times$240  &  22.8  &  19.4  &  14.8  &  M4  \\
\enddata

\tablenotetext{a}{Faintest magnitude that has a S/N of 5.}

\tablenotetext{b}{Faintest magnitude that has a S/N of S/N$_{\rm floor}$~=~100.}

\tablenotetext{c}{Saturation limits derived for a Gaussian PSF with a
  FWHM of 0\farcs8, a gain of 1 $e^-$ DN$^{-1}$, a full-well depth of
  3.0$\times$10$^4$~$e^-$.}

\tablenotetext{d}{Earliest spectral type that is not saturated at
  100~pc.}

\end{deluxetable}

\end{document}